\documentclass[aps, twocolumn, superscriptaddress, showpacs, nofootinbib]{revtex4-1}
\usepackage{graphicx}
\usepackage{bm}
\usepackage{amsmath, amsthm,amssymb,url}
\usepackage[normalem]{ulem}
\usepackage{color}
\usepackage{array}
\usepackage{float}
\usepackage{subcaption}
\RequirePackage{lineno}
\usepackage{graphicx}
\usepackage{caption}
\usepackage{subcaption}
\usepackage{draftwatermark}
\SetWatermarkText{}
\SetWatermarkScale{5}
\def\Msun{{\rm M}_{\odot}}

\def\be{\begin{equation}}
\def\ee{\end{equation}}
\def\bea{\begin{eqnarray}}
\def\eea{\end{eqnarray}}
\def\bn{\begin{enumerate}}
\def\en{\end{enumerate}}

\def\c{\times}

\usepackage{lineno}
\usepackage{color}
\usepackage{xparse,graphicx,xcolor}

\makeatletter

\newcommand{\Rmnum}[1]{\expandafter\@slowromancap\romannumeral #1@}
\makeatother

\begin{document}

\captionsetup[figure]{labelfont={bf},labelformat={default},labelsep=period,name={Figure.}}
\preprint{------}

\title{Search for intermediate mass black hole binaries in the first and second observing runs of the Advanced LIGO and Virgo network}

\author{B.~P.~Abbott}
\affiliation{LIGO, California Institute of Technology, Pasadena, CA 91125, USA}
\author{R.~Abbott}
\affiliation{LIGO, California Institute of Technology, Pasadena, CA 91125, USA}
\author{T.~D.~Abbott}
\affiliation{Louisiana State University, Baton Rouge, LA 70803, USA}
\author{S.~Abraham}
\affiliation{Inter-University Centre for Astronomy and Astrophysics, Pune 411007, India}
\author{F.~Acernese}
\affiliation{Dipartimento di Farmacia, Universit\`a di Salerno, I-84084 Fisciano, Salerno, Italy}
\affiliation{INFN, Sezione di Napoli, Complesso Universitario di Monte S.Angelo, I-80126 Napoli, Italy}
\author{K.~Ackley}
\affiliation{OzGrav, School of Physics \& Astronomy, Monash University, Clayton 3800, Victoria, Australia}
\author{A.~Adams}
\affiliation{Christopher Newport University, Newport News, VA 23606, USA}

\author{C.~Adams}
\affiliation{LIGO Livingston Observatory, Livingston, LA 70754, USA}
\author{R.~X.~Adhikari}
\affiliation{LIGO, California Institute of Technology, Pasadena, CA 91125, USA}
\author{V.~B.~Adya}
\affiliation{OzGrav, Australian National University, Canberra, Australian Capital Territory 0200, Australia}
% ...
\collaboration{The LIGO Scientific Collaboration and the Virgo Collaboration}


\date{\today}
\begin{abstract}


	Gravitational wave astronomy has been firmly established with the detection of gravitational waves from the merger of ten stellar mass binary black holes and a neutron star binary.
	This paper reports on the all-sky search for gravitational waves from intermediate mass black hole binaries in the first and second observing runs of the Advanced LIGO and Virgo network.
	The search uses three independent algorithms:
	two based on matched filtering of the data with waveform templates of gravitational wave signals from compact binaries, and a third, model-independent algorithm that employs no signal model for the incoming signal.
	No intermediate mass black hole binary event was detected in this search.
	Consequently, we place upper limits on the merger rate density for a family of intermediate mass black hole binaries. In particular, we choose sources with total masses $M=m_1+m_2\in[120,800]$~M$_\odot$ and mass ratios $q = m_2/m_1 \in[0.1,1.0]$.
	For the first time, this calculation is done using numerical relativity waveforms (which include higher modes) as models of the real emitted signal.
	We place a most stringent upper limit of {$0.20$~Gpc$^{-3}$yr$^{-1}$} (in co-moving units at the 90\% confidence level) for equal-mass binaries with individual masses $m_{1,2}=100$~M$_\odot$ and dimensionless spins $\chi_{1,2}= 0.8$ aligned with the orbital angular momentum of the binary.
	This improves by a factor of {$\sim 5$}  that reported after Advanced LIGO's first observing run.

\end{abstract}


\maketitle

\section{Introduction}
\label{SecI}
	The first two observing runs of Advanced LIGO and Virgo (O1 and O2 respectively) have significantly enhanced our understanding of black hole (BH) binaries in the universe.
	Gravitational waves (GWs) from 10 binary black hole mergers with total mass between $18.6^{+3.1}_{-0.7}~{\rm M}_\odot$ and $85.1^{+15.6}_{-10.9}~{\rm M}_\odot$ were detected during these two observing runs  
\cite{GW150914-DETECTION,GW151226-DETECTION,GW170104-DETECTION,GW170608-DETECTION,GW170817-DETECTION,GW170814-DETECTION,TheLIGOScientific:2016pea,O1-O2-catalog-paper}. %
	These observations have revealed a new population of heavy stellar mass BH components of up to 50~M$_\odot$, for which we had no earlier electromagnetic observational evidence~\cite{O1-O2-catalog-paper,O1-O2-rate_paper}.
	This finding limit is consistent with the formation of heavier BHs from core collapse being prevented by a mechanism known as \textit{pulsational pair-instability supernovae} (PISN)~\cite{Spera:2017fyx,Woosley:2016hmi,Giacobbo:2017qhh,Belczynski:2016jno}. 
According to this {idea}, stars with {helium} core mass in the range $\sim 32 - 64 M_\odot$ undergo pulsational pair instability leaving behind remnants of $\lesssim{} 65 M_\odot$. Stars with {helium} core mass in the range $\sim 64 - 135 M_\odot$ undergo pair instability and leave no remnant, while stars with {helium} mass $\gtrsim{}135 M_{\odot}$ are thought to directly collapse to intermediate-mass black holes.

Intermediate mass black holes (IMBHs) are BHs heavier than stellar mass BHs but lighter than supermassive black holes (SMBHs), which places them roughly in the range of {$10^2-10^5~{\rm M}_\odot$} \cite{Merzcua:2017,2017mbhe.confE..51K}.
Currently there is only {indirect observational evidence}.
{Observations} include probing the mass of the central BH in galaxies as well as massive star clusters with direct kinematical measurements 
which has led to recent claims for the presence of IMBHs \cite{Kiziltan:imbh,denBrok,AtakanGurkan:2003hm}.
{Other observations come from} the extrapolation of several scaling relations between the mass of the central SMBH and their host galaxies \cite{Kormendy:2013dxa} to the mass range of globular clusters \cite{Graham:2012bs,2013ApJ...764..151G}.
In this way, several clusters have been found to be good candidates for {having} IMBHs in their centers \cite{vandenBosch:2005uk,vanderMarel:2009mc,Anderson:2009ku}.
If present, IMBHs would heat up the {cores} of these clusters, strongly influencing the distribution of the stars in the cluster and their dynamics, leaving a characteristic imprint in the surface brightness profile, as well as {in the} mass-to-light ratio \cite{Baumgart:2017imbh}. 
Controversy exists regarding the interpretation of these observations, as some of them can also be explained by a high concentration of stellar-mass BHs or the presence of binaries\cite{Baumgardt:2002rc,vandenBosch:2005uk,vanderMarel:2009mc,Anderson:2009ku}.
Empirical mass scaling relations of quasi-periodic oscillations \cite{doi:10.1146/annurev.astro.44.051905.092532} in luminous X-Ray sources have also provided evidence for IMBH \cite{Pasham:2015tca}.
Finally, IMBHs have been proposed as candidates to explain ultra-luminous X-Ray (ULX) sources in nearby galaxies, which are brighter than the accreting X-ray sources with stellar mass BHs \cite{Kaaret:2017tcn,Farrell:2010bf}.
However, neutron stars or stellar-mass black holes emitting above their Eddington luminosity could also account for such observations.
In summary, no definitive evidence of IMBHs has yet been obtained.

The possible astrophysical formation channels of IMBHs remain uncertain.
Proposed channels include the direct collapse of massive first generation, low metallicity Population III stars~\cite{Madau:2001sc,Schneider:2001bu,Ryu:2016dou,2014MNRAS.442.2963K} and mergers of stellar mass BHs in globular clusters~\cite{PortegiesZwart:2004ggg} and multiple collisions of stars in dense young star clusters \cite{Miller:2001ez, PortegiesZwart:2002iks,AtakanGurkan:2003hm,Mapelli:2016vca, Leigh:2014oda}, among others~\cite{Marchant:2016wow}.
Further, some astrophysical scenarios \cite{Merzcua:2017} indicate that SMBHs in galactic centers might be formed from hierarchical mergers of IMBHs~\cite{2017mbhe.confE..51K,King:2004ri}. The direct observation of IMBHs with gravitational waves could strengthen the possible evolutionary link between stellar mass BHs and SMBHs. Finally, observing an IMBH population would help to understand details  of the pulsational pair-instability supernovae mechanism.

The GW observation of a coalescing binary consisting of at least one IMBH component or resulting in an IMBH remnant, which we will term an IMBHB, could provide the first definitive confirmation of the existence of IMBHs.
In fact, IMBHBs are the sources that would emit the most gravitational-wave energy in the LIGO-Virgo frequency band, potentially making them detectable to distances (and redshifts) beyond that of any other LIGO-Virgo source~\cite{Abbott:2017iws}.
Even in the absence of a detection, a search for IMBHBs provides stringent constraints on their merger rate density, which has implications for potential IMBHB and SMBH formation channels.

IMBHs are not only interesting from an astrophysical point of view, they are also excellent laboratories to test general relativity in the strong field regime \cite{TheLIGOScientific:2016src,Yunes:2016jcc,Yunes:2013dva,Gair:2012nm}.
Their large masses would lead to strong merger and ringdown signals in the Advanced LIGO-Virgo frequency band.
Therefore, higher modes might be visible in IMBHB signals because those modes are especially strong in the merger and ringdown stages.
The observation of multimodal merger and ringdown signals is paramount to understanding fundamental properties of general relativity, such as the no-hair theorem~\cite{Kamaretsos:2011um,Meidam:2014jpa,Thrane:2017lqn,Carullo:2018sfu} and BH kick measurements~\cite{Gonzalez:2006md,Campanelli:2007cga,CalderonBustillo:2018zuq}.

The first search for GWs from IMBHBs was carried out in the data from initial LIGO and initial Virgo (2005-2010) ~\cite{Virgo:2012aa, Aasi:2014iwa}. Owing to {the} large masses of IMBHBs, {such systems are} expected to merge at low frequencies where the {initial detectors were} less sensitive due to the presence of several noise sources, such as suspension noise, thermal noise and optical cavity control noise. As a result, the those detectors were sensitive to only {the}  merger and ringdown phases of the IMBHB systems. Initial IMBHB {analyses applied either} the model-independent time-frequency searches \citep{klimenko:2005xv} or ringdown searches. No IMBHB merger was detected in these searches.

Because of the improved low-frequency sensitivities of the Advanced LIGO and Advanced Virgo detectors~\cite{TheLIGOScientific:2014jea,TheVirgo:2014hva}, IMBHB signals {are visible} in band for a longer period of time, which increases the effectiveness of modeled searches that {use} more than just the ringdown portion of an IMBHB's waveform.
Ref.~\cite{Abbott:2017iws} reports results from a combined search for IMBHBs that used two independent search algorithms: a matched-filter analysis, called GstLAL~\cite{Cannon:2011vi,Privitera:2013xza,Messick:2016aqy}, {which uses} the inspiral, merger, and ringdown portions of the IMBHB waveform and the model-independent analysis coherent WaveBurst (cWB) \citep{klimenko:2005xv}.
No IMBHBs were found by these searches, and upper limits on the merger rate density for 12 targeted IMBHB sources with total mass between $120~{\rm M}_\odot-600~{\rm M}_\odot$ and mass ratios down to $1/10$ were obtained.
The most stringent upper limit on the merger rate density from this combined analysis was $0.93~{\rm {\rm Gpc}}^{-3}~{\rm {\rm yr}}^{-1}$ for binaries consisting of two $100~{\rm M}_\odot$ BHs with dimensionless spin magnitude $0.8$ aligned with the system's orbital angular momentum.

All upper limits on the IMBHB merger rate reported in past searches \cite{Virgo:2012aa,Aasi:2014iwa,Abbott:2017iws} were obtained using models for the GW signal that {include only} the dominant radiating mode, namely $(\ell,m)=(2,\pm2)$, of the GW emission \cite{Taracchini:2012ig}.
However, it has been shown that higher modes contribute more substantially to signals emitted by heavy binaries. 
This impact increases as the system becomes more asymmetric in mass~\cite{Pekowsky:2012sr, Varma:2014jxa}, as the spin of the BHs becomes more negative~\cite{Bustillo:2015qty, Varma:2016dnf}, and as the precession in the binary becomes stronger~\cite{Bustillo:2016gid}. 
As a consequence, the omission of higher modes 
leads in general to more conservative upper limits on the IMBHB merger rate \cite{CalderonBustillo:2017skv}. 
In this work, we improve on past studies in two distinct ways. We use numerical relativity (NR) simulations with higher modes to model GW signals from IMBHBs for computing upper limit estimates.
Additionally, our combined analysis now includes the matched-filter search PyCBC~\cite{Canton:2014ena,Usman:2015kfa} in addition to GstLAL and cWB. 
Because of these novelties, we have, in addition to analyzing the O2 data set, reanalyzed the O1 data set and report {here} combined upper limits for the O1 and O2 observing runs.
In this paper, we report upper limits on the merger rate density of 17 targeted (non-precessing) IMBHB sources. 
Our most stringent upper limit {is} {$0.20 ~{\rm Gpc}^{-3}~{\rm yr}^{-1}$} for equal-mass binaries with component spins aligned with the orbital angular momentum of the system and dimensionless magnitudes $\chi_{1,2} = 0.8$.

The rest of this paper is organized as follows. In Sec.~\ref{sec:SecII} we describe the data set, outline the individual search algorithms that make up the combined search, and report our search's null detection of IMBHBs.
In Sec.~\ref{sec:SecIII}  we describe the NR simulations that we use to compute upper limits on IMBHB merger rates report these for the case of 17 IMBHB sources.
We draw final conclusions in Sec.~\ref{sec:conclusion}.

\section {IMBHB search in O1 and O2 data}
\label{sec:SecII}


\subsection{Data Summary}

This analysis was carried out using O1 and O2 data sets from the two LIGO (Livingston and Hanford) detectors and Virgo. 
We have used the final calibration, which was produced after the conclusion of the run, {including} compensation for frequency-dependent fluctuations in the calibration \cite{Cahillane:2017vkb,Viets:2017yvy,Acernese:2018bfl}.
Well identified sources of noise have also been subtracted from the strain data as explained in Refs.~\cite{Driggers:2018gii,Acernese:2018bfl}.
The maximum calibration uncertainty across the frequency band of [10-5,000] Hz for the two LIGO detectors is $\sim 10\%$ in amplitude and $\sim 5$ degrees in phase for O1 and $\sim 4\%$ in amplitude and $\sim 3$ degrees in phase for O2 \cite{TheLIGOScientific:2016pea,Cahillane:2017vkb}.
For Virgo we consider an uncertainty of $5.1\%$ in amplitude and $2.3$ degrees in phase \cite{Acernese:2018bfl}.  
After removing data with significant instrumental disturbance,
we use $48.6$ days and $118.0$ days of joint Hanford-Livingston data from the O1 and O2 observing runs respectively. 
The Virgo detector joined the LIGO detectors during the last $\sim 15$ days of O2, which provided with an additional {$4.0~$} days of coincident data with either of Hanford-Virgo or Livingston-Virgo network. 
The data from O1 and O2 was divided into 9 and 21 blocks respectively with coincident time ranging from $4.7-7.0$~days.
For more details, see Ref.~\cite{O1-O2-catalog-paper}.
	
	\subsection{Search algorithms}
	\label{Sec:Search}

	We combine the two matched-filter searches, namely GstLAL~\cite{sachdev2019gstlal,Cannon:2011vi,Privitera:2013xza,Messick:2016aqy} and PyCBC \cite{Canton:2014ena,Usman:2015kfa}, and one model-independent analysis, cWB \cite{Klimenko:2015ypf}, into a single IMBHB search.
	The two model-based matched filtering analyses use a bank of templates made of pre-computed compact binary merger GW waveforms.
	Matched filter based analyses are optimal to extract known signals from stationary, Gaussian noise \cite{MatchedFilter}.
	However, the templates we use are limited to non-eccentric, aligned-spin systems.
	They contain {only} the dominant waveform mode of the GW emission and omit higher modes \cite{Varma:2014jxa,Capano:2013raa}.
	Additionally, Advanced LIGO and Virgo data are known to contain a large number of short noise transients \cite{Cabero:2019orq}, which can mimic short GW signals like those emitted by IMBHBs.
	While matched-filter searches use several techniques to discriminate between noisy transients and real GW events \cite{Canton:2013joa,Nitz:2017lco,Messick:2016aqy}, they are known to {lose significant} efficiency when looking for short signals like those from IMBHBs.
	Therefore, the IMBHB search is carried out jointly with an analysis that can identify short-duration GW signals without a model for the morphology of the GW waveform.
	In this search, all three analyses use O1 and O2 Advanced LIGO data.
	However, {because of} the incomparable sensitivities between the Advanced LIGO detector and Advanced Virgo detector, only the GstLAL analysis uses Virgo data, as is done in Ref.~\cite{O1-O2-catalog-paper}.

	\subsubsection{Modeled analyses}

	The matched-filter analyses GstLAL and PyCBC use templates that span the parameter space of neutron stars, stellar-mass BHs, and IMBHs.
	In this study, we use the same two searches reported in Ref.~\cite{O1-O2-catalog-paper} to calculate upper limits on the merger rate density of IMBHBs.

	The matched-filter signal-to-noise ratio (SNR) time series is computed for every template.
	Triggers are produced when the SNR time series surpasses a predetermined threshold, where clusters of triggers are trimmed by maximizing the SNR within small time windows.
	In addition, a signal consistency veto \cite{Allen:2004gu,Nitz:2017lco,Messick:2016aqy} is calculated for each trigger. 
	A list of GW candidates is constructed from triggers generated by common templates that are coincident in time across more than one detector, where the coincidence window takes into account the travel time between detectors.
	Next, a ranking statistic is calculated for each candidate that estimates a likelihood ratio that the candidate would be observed in the presence of a GW compared to {a pure-noise expectation}.
	Finally, a $p$-value\footnote{The probability that noise would produce a trigger at least as significant as the observed candidate.} $P$ is determined by comparing the value of its ranking statistic to that of triggers coming from background noise in the data. 
	A detailed description of the GstLAL and PyCBC pipelines can be found in Refs.~\cite{sachdev2019gstlal,Cannon:2011vi,Privitera:2013xza,Messick:2016aqy} and \cite{Canton:2014ena,Usman:2015kfa}, respectively; additionally, details outlining how candidates are ranked across observing runs can be found in Ref.~\cite{O1-O2-catalog-paper}.

	The GstLAL analysis uses the template bank described in Ref.~\cite{gstlal_bank}.
	The region of this bank that overlaps the IMBHB parameter space, which starts at a total mass of 100~M$_\odot$, reaches up to a total mass of 400~M$_\odot$ in the detector frame and covers mass ratios in the range of $1/98<q<1$.
	The waveforms used are a reduced-order-model of the SEOBNRv4 approximant~\cite{Bohe:2016gbl}.
	The spin of these templates are either aligned or anti-aligned with the orbital angular momentum of the system with dimensionless magnitudes less than 0.999.

	The PyCBC analysis uses the template bank described in Ref.~\cite{DalCanton:2017ala}.
	The region of this bank that overlaps the IMBHB parameter space reaches up to a total mass of 500~M$_\odot$ in the detector frame, excluding templates with duration below 0.15~s, and covers the range of $1/98<q<1$.
	The waveforms used are also a reduced-order-model of the SEOBNRv4 approximant, and the aligned or anti-aligned dimensionless spin magnitudes are less than 0.998.

	\subsubsection{Un-modeled analysis}
	\label{sec:unmodeled}

Coherent WaveBurst (cWB) is the GW transient detection algorithm designed to look for unmodeled short-duration GW transients in the multi-detector data from interferometric GW detector networks. Designed to operate without a specific waveform model,
cWB identifies coincident excess power in the wavelet time-frequency representations
of the detector strain data~\cite{Necula:2012zz}, for signal frequencies
up to 1 kHz and durations up to a few seconds. The search identifies events that are coherent
in multiple detectors and reconstructs the source sky location and signal waveforms
by using the constrained maximum likelihood method~\cite{Klimenko:2015ypf}.
The cWB detection statistic is based on the coherent energy $E_{\rm c}$ obtained by cross-correlating
the signal waveforms reconstructed in multiple detectors. It is proportional to the
network SNR and used to rank the events found by cWB.

To improve the robustness of the algorithm against non-stationary detector
noise, cWB uses signal-independent vetoes, which reduce the high rate of the initial excess power triggers.
The primary veto cut is on the network correlation coefficient $c_{\rm c} = E_{\rm c}/(E_{\rm c}+E_{\rm n})$,
where $E_{\rm n}$ is the residual noise energy estimated after the reconstructed signal is subtracted from the data.
Typically, for a GW signal $c_{\rm c} \approx 1$ and for instrumental glitches $c_{\rm c} \ll 1$.
Therefore, candidate events with $c_{\rm c} < 0.7$ are rejected as potential glitches.

To improve the detection efficiency of IMBHBs as well as to reduce the false alarm rates (FARs), the cWB analysis employs additional selection cuts based on the nature of IMBHB signals.
IMBHB signals have two distinct features in the time-frequency representation.
First, the signal frequencies {lie} below 250~Hz.
We use this to exclude all the non-IMBHB events in the search, including noise events.
Secondly, the inspiral signal duration in the detector band is relatively short, which leads to relatively low SNR in the inspiral phase as compared to the merger and ringdown phases. 
In the cWB framework, chirp mass (${\cal{M}}=(m_1~m_2)^{3/5} ~ M^{-1/5}$) is estimated using the frequency evolution of a signal's inspiral.
However, in the case of low SNRs, we cannot accurately estimate the chirp mass of the binary \cite{Tiwari:2015gal}; still, we use this framework to introduce additional cuts on the estimated chirp mass  to reject non-IMBHB signals. The simulation studies show that IMBHB signals are recovered with $|{\cal{M}}|>10 ~\Msun$ which we use in this search\footnote{Negative $\cal{M}$ values correspond to frequencies decreasing with time, which could be due to the pixels corresponding to ringdown part.}. 
We apply this selection cut to reduce the noise background when producing the candidate events.

For estimation of the statistical significance of the candidate event, each event is ranked against a sample of background triggers obtained by repeating the analysis on time-shifted data \cite{GW150914-DETECTION}. To exclude astrophysical events from the background sample, the
time shifts are selected to be much {longer} (1 second or more) than the expected signal time delay between the detectors. By using different time shifts, a sample of background events
equivalent to approximately 500 years of background data is accumulated for each
of the  30 blocks of data. The cWB candidate {events that} survived the cWB
selection criteria, are assigned a FAR given by the rate of
the corresponding background events with the coherent network SNR value larger than that of the candidate event.

	\subsection{Combined search}
	\label{sec:combined}

Each of our three algorithms produces its own list of GW candidates, characterized by GPS time, FAR and associated $p$-value $P$.
These three lists are then combined into a common single list of candidates.
To avoid counting candidates more than once, candidates within a time window of 100~ms across different lists are assumed to be the same.
To account for the use of three search algorithms, we apply a conservative trials factor of 3 and assign each candidate a new $p$-value given by
	\begin{equation}
	\label{pbar}
	\bar{P}=1-(1-P_{\rm min})^3, \
	\end{equation}
where $P_{\rm min}$ denotes the minimum $p$-value reported across the pipelines.
This is equivalent to assuming that the three searches produce independent lists of candidates.\footnote{In general, correlations between searches would lead to a trials factor less than 3. 
However, at the time, we are not able to quantify this, and we choose to adopt the most conservative approach.}
We note that while this choice of trials factor affects the significance of individual triggers, it will not change the numerical value of our upper limits.
See Appendix~\ref{sec:pval} for a more detailed discussion.

	\begin{table} [htb!]
		\setlength\extrarowheight{5pt}
		\begin{center}
			\begin{tabular}{ |p{.5 cm} |p{1.7 cm} | p{1.5 cm} | p{ 1.5 cm} | p{1.0 cm} | p{1.0 cm} | p{0.8 cm} | p{0.8 cm} ||}
				\hline
				\bf{No} & {\bf{Date}}    &   {\bf{UTC time}}   &   {\bf{Pipeline}} &  {\bf{FAR (${\rm yr}^{-1}$)}}   &   {\bf{SNR}} & {\bf{$\boldsymbol{{P_{\rm min}}}$}}  \\  \hline  
				
				1 &2017-05-02 & {04:08:44.9}  &            cWB    &       0.34   & 11.6 & 0.14 \\
				2 & 2017-06-16 & 19:47:20.8 & 	  PyCBC	 &       1.94	& 9.1 & {0.59} \\
				3 & 2015-11-26 & {04:11:02.7} &             cWB    &      2.56    &  7.5 & 0.68\\
				4 &2017-06-08 & {23:50:52.3} &            cWB    &      3.57    & 10.0  & 0.79\\
				5 & 2017-04-05 & 11:04:52.7 &	          GstLAL &	4.55	& 9.3  & 0.88 \\
				6 & 2015-11-16 & 22:41:48.7 &	          PyCBC &	4.77	& 9.0  & {0.88} \\
				7 &2016-12-02 & 03:53:44.9 &	          GstLAL &	6.00	& 10.5  & 0.94\\
				8 & 2017-02-19 & 14:04:09.0 &	          GstLAL &	6.26	& 9.6  & 0.95 \\
				9 & 2017-04-23 & 12:10:45.0 &	  GstLAL &	6.47	& 8.9 & 0.95\\
				10 & 2017-04-12 & 15:56:39.0 &	  GstLAL &	8.22	& 9.7 &0.98 \\ \hline 
			\end{tabular}
		
		\caption{Details of the ten most significant events (excluding all published lower mass events). We report the date, UTC time, observing pipeline (individual analysis that observed the event with the highest significance), FAR, SNR, and $P_{\rm min}$ for each event.  The combined $p$-value $\bar{P}$ of each event is calculated using Eq.~\ref{pbar}. In the table, the events are tabulated in increasing value of $P_{\rm min}$.}
		\label{tab:tab1}
	\end{center}
	\end{table}

	\subsection{Search results}
	
	\label{sec:SecIV}

Here we report results from the combined cWB-GstLAL-PyCBC IMBHB search on full O1-O2 data. 
The top 21 most significant events from the combined search include the 11 GW events {published} in Ref.~\cite{O1-O2-catalog-paper}, namely GW150914 \cite{GW150914-DETECTION}, GW151012, GW151226 \cite{GW151226-DETECTION}, GW170104 \cite{GW170104-DETECTION}, GW170608 \cite{GW170608-DETECTION}, GW170729, GW170809, GW170814 \cite{GW170814-DETECTION}, GW170817 \cite{GW170817-DETECTION}, {GW170818}, and GW170823, and 10 events tabulated in Table~\ref{tab:tab1}. 
All the events in Table~\ref{tab:tab1} have a FAR much larger than any of the GWs reported in Ref.~\cite{O1-O2-catalog-paper}; no event in this list was found with enough significance to claim an IMBHB detection.

The top-ranked event\footnote{We note that the most significant event in the O1 search reported in Ref.~\cite{Abbott:2017iws} is the third event in this Table.} from Table~\ref{tab:tab1} was observed by cWB in O2 data on May 2, 2017 at 04:08:44 UTC with a combined SNR of 11.6 in the two Advanced LIGO detectors and a significance of $P_{\rm cWB}=P_{\rm min}=0.14$. 
Applying eq.~\ref{pbar}, this event has a combined $p$-value of $\bar{P}=0.36$, too low to claim it as a gravitational wave detection.

Despite the low significance of this trigger, its characteristics were consistent with those of an IMBHB, and we decided to perform detailed data quality and parameter estimation follow-ups.\footnote{See Appendix~\ref{sec:pe} for further details regarding the parameter estimation investigations of this candidate.}
In order to check for the presence of environmental or instrumental noise, this event was vetted with the same procedure applied to triggers of marginal significance found in previous searches~\cite{O1-O2-catalog-paper} in O1-O2 data.
These checks identified a correlation between the trigger time and the glitching of optical lever lasers at the end of one arm of the Hanford detector.
This is a known instrumental artifact previously observed to impact GW searches \cite{nuttall_detchar,TheLIGOScientific:2017lwt}.
The time of this trigger was not discarded by the pre-tuned data quality veto designed to mitigate the effects of these optical lever laser glitches.
However, these vetoes are tuned for high efficiency and minimal impact on analyzable time rather than exhaustively removing all non-Gaussian features in the data.
Further follow-up indicates that this instrumental artifact is likely contributing power to the gravitational wave strain channel at the time and frequency of the trigger.
Given the SNR of the purported signal in the Hanford detector and the relatively low significance of the reported false alarm rate, we conclude that this trigger is likely explained by detector noise.

\section{Upper limits on merger rates}
	\label{sec:SecIII}
	Given that no IMBHB signal was detected by our search, we proceed to place upper limits on the coalescence rate of these objects.
	This is done by estimating the sensitivity of our search to an astrophysically motivated population of simulated IMBHB signals that we inject in our detector data.
	However, given the absence of  well motivated population estimates of IMBHBs, we opt for sampling the parameter space in a discrete manner (for details, please see Appendix \ref{sec:appeB}).
	As a consequence, in this section we estimate the sensitive distance reach as well as the upper limit on merger rate density for {17} selected fiducial IMBHB sources, tabulated in Table~\ref{tab:tab23} of Appendix~\ref{sec:appeB} using the loudest event method~\cite{Biswas:2007ni}, following the procedure outlined in Ref.~\cite{Abbott:2017iws} and described again in Appendix~\ref{sec:sensivetive_distance_reach}. 
	For a given IMBHB source, 
	gravitational waveforms from simulated systems scattered through space are injected into the data and recovered by each of the three analyses.
	In this section, we describe our simulation set and present our findings.

	\subsection{Injection set}
	\label{simulationset}

	\newcolumntype{C}[1]{>{\centering\arraybackslash}m{#1}}

	Ref.~\cite{Abbott:2017iws} reports upper limits on the merger rate density for 12 IMBHB systems in its Table I. 
	The waveform simulations used to compute upper limits in that study  {contain only} the dominant quadrupolar mode of the GW emission. 
	In this work, we use highly accurate NR simulations computed by the SXS~\cite{Mroue:2013xna}, RIT~\cite{Healy:2017psd}, and GeorgiaTech~\cite{Jani:2016wkt} codes, which include higher modes.
	Since higher modes are particularly important for large asymmetries in mass and for high total mass binaries, in this study we extend our parameter space to mass ratios as low as $q=1/10$ and total masses as high as $M=800{\rm M}_\odot$ (see Appendix \ref{sec:appeB} Table~\ref{tab:tab23} for a detailed list). 
	In general, NR simulations can include modes of arbitrary $(\ell,m)$ for a given set of masses in the parameter space.
	However, weak modes are sometimes dominated by numerical noise and do not agree when {compared} across different numerical codes. In fact, we disregard numerical modes with $\ell \ge 5$, because these have comparatively small amplitudes.  The $\ell\ge 5$ modes with $m=\pm \ell$ have also particularly short wavelengths, which makes it more challenging for numerical relativity codes to resolve the propagation of these modes away from the binary.
	In order to assess the accuracy of the remaining modes, we only choose IMBHB simulations for which higher modes  have been computed by at least two different NR codes. 
	We select only those higher modes that agree to an overlap of at least 0.97 across all available NR codes for each of 17 the selected simulations.\footnote{We did allow for overlaps below 0.97 if the mode's contribution to the waveform was negligible.}

	The higher modes that passed this criteria and were included in our analysis were the following: $(\ell,m)=\{(2,\pm1),(2,\pm2),(3,\pm2),(3,\pm3),(4,\pm2),(4,\pm3),(4,\pm4)\}$.
	Notably, the $(2,2)$ mode agrees across NR codes to an overlap $>0.995$ for every IMBHB source considered in this study; the two modes closest to the 0.97 overlap threshold were the $(4,4)$ and $(4,3)$ modes.
	We note that, similar to what was described in Ref.~\cite{CalderonBustillo:2017skv}, omission of $\ell > 4$ modes may lead to an underestimation of the  power within the detector band radiated by the largest mass binary BHs.
	
	Of the 17 selected sources, we include three cases with spins aligned or anti-aligned with the total angular momentum of the binary with dimensionless spin magnitudes $|\chi_{1,2}|=0.8$.
	The IMBHB injections are uniformly distributed in the binary orientation parameters ($\varphi$,$\cos(\iota)$), uniformly distributed in co-moving volume up to red shift $z \sim 1 $ (luminosity distance of $6.7~{\rm Gpc}$) using the TT+lowP+lensing+ext cosmological parameters given in Table IV of Ref.~\cite{Ade:2015xua}, and individually redshifted according to this cosmological model. {The overall effect of cosmological redshift is to shift each GW signal to lower frequencies.  At a given redshift, the mass of the injection in the detector frame is (1+z) times larger than the source frame mass, and the luminosity distance is (1+z) 
		times the comoving distance. At redshifts of z=1, this results in a decrease in SNR, ranging from $\sim 20\%$ for an equal-mass $M=100 ~\Msun$ face-on system to $\sim 50\%$ for an equal-mass, $M=200 ~\Msun$ face-on system.
		} The injections are spaced roughly uniformly in time with an interval of at least 80~seconds over the {$T_{0} =413.71$} days of O1-O2 observing time,
	and each injection set covers a total space-time volume $\langle VT\rangle_{\rm tot}= 110.68 \; {\rm Gpc}^3 \; {\rm yr}$.

	\subsection{Sensitive distance reach and merger rate density estimate}
	\label{sec:SensDist}
 
	We use the loudest event method~\cite{Biswas:2007ni} to calculate the sensitive distance reach of our search and to place upper limits on the merger rate density of IMBHBs (see Appendix \ref{sec:appeB} for a detailed description of our procedure).
	The results of our combined search are reported using a combined $p$-value of $\bar{P}=0.36$ given by that of our loudest event in Table~\ref{tab:tab1}.

\begin{figure*}[!htp]
	\hspace{-3.5 cm}
	\centering
	\begin{minipage}[b]{.45\textwidth}
		\includegraphics[scale=0.55]{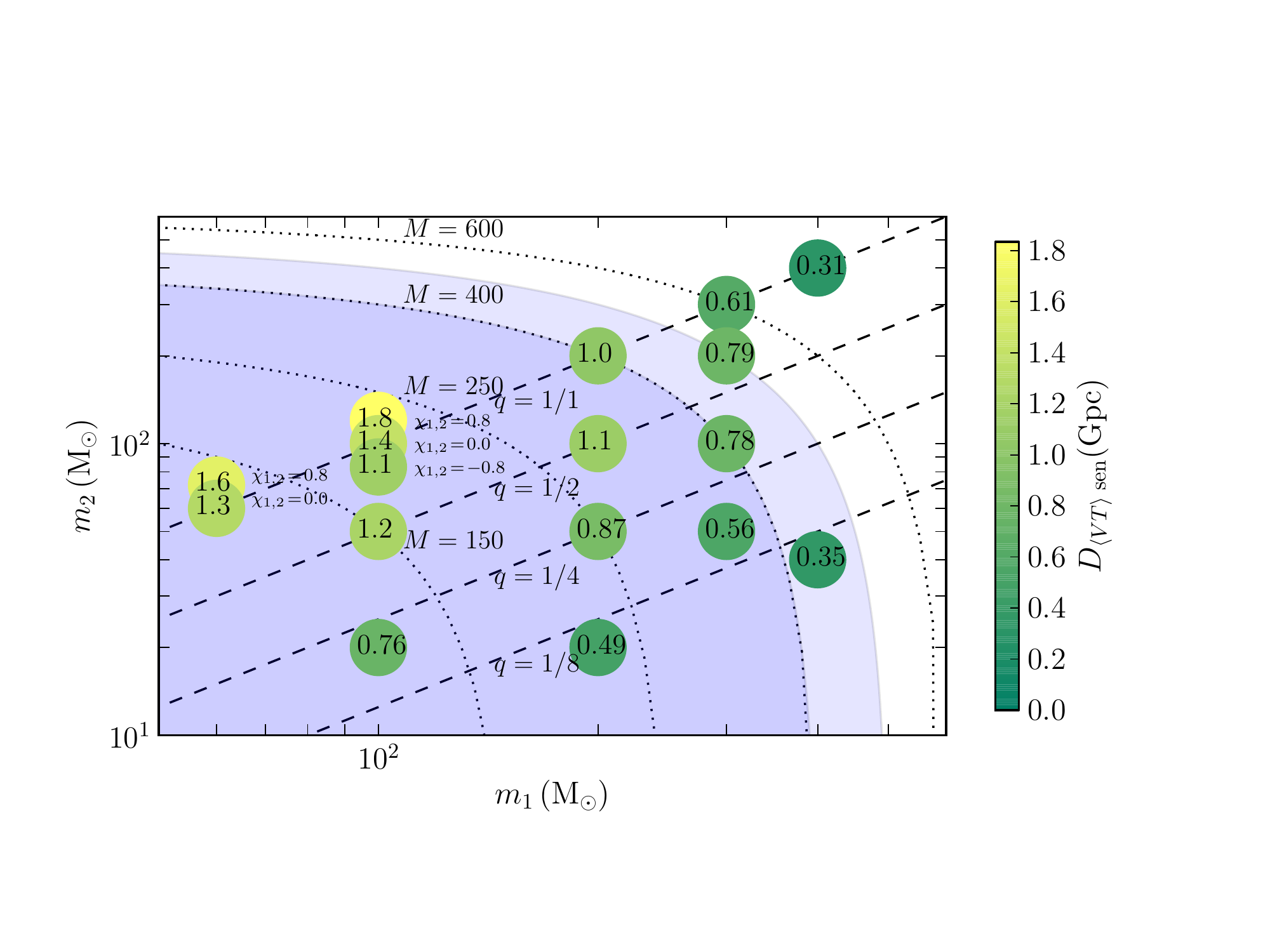}
	\end{minipage}
	\qquad \qquad \qquad
	\begin{minipage}[b]{.45\textwidth}
		\includegraphics[scale=0.55]{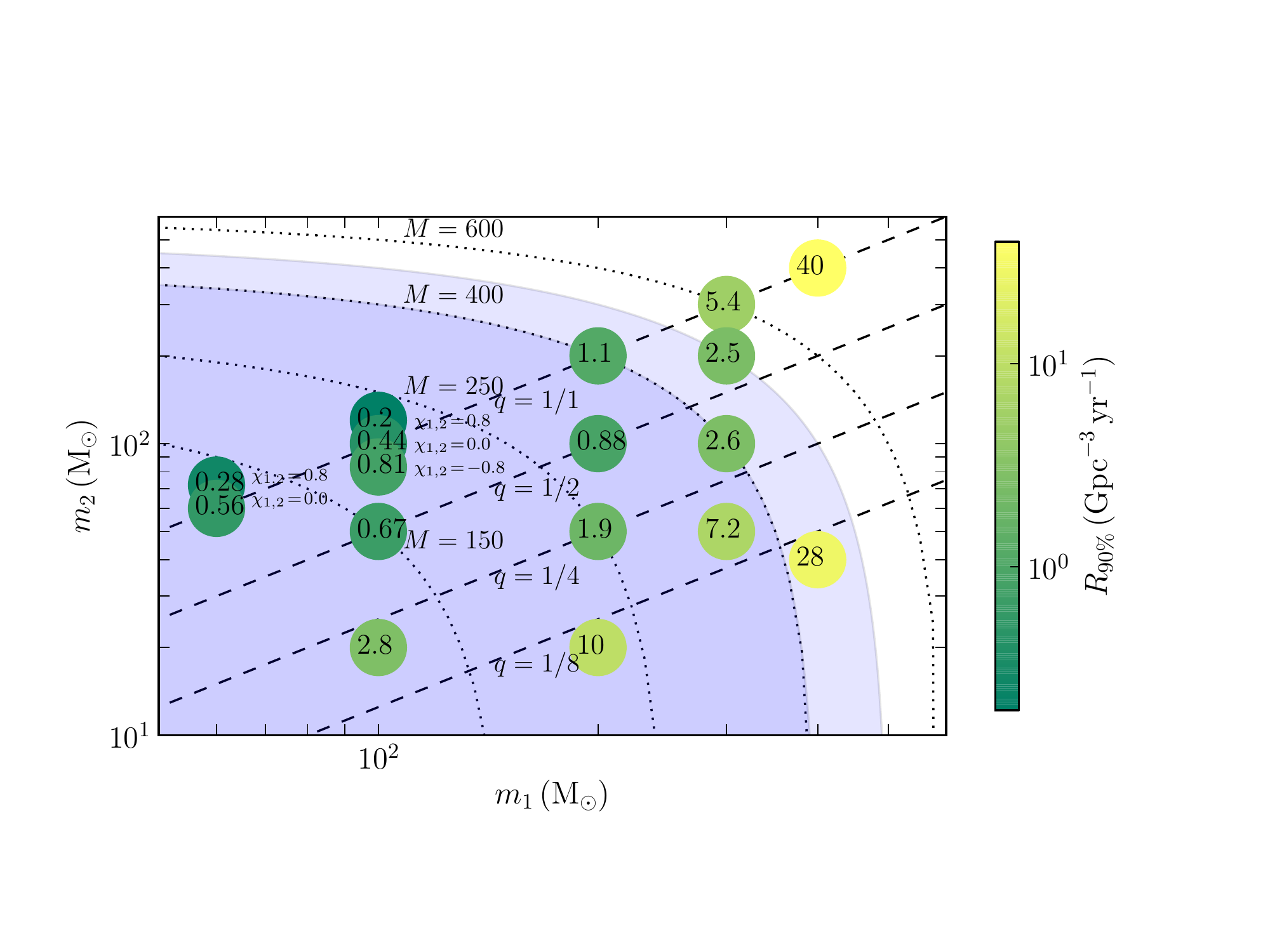}
	\end{minipage}
	\caption{The sensitive distance reach ($D_{\langle VT \rangle_{\rm sen}}$) in ${\rm Gpc}$ (left-panel) and $90\%$ upper limit on merger rate density ($R_{90\%}$) in ${\rm Gpc}^{-3}~{\rm yr}^{-1}$ (right-panel) for the 17 targeted IMBHB sources in the $m_1-m_2$ plane. Each circle corresponds to one class of IMBHBs in the source frame with a number in the circle indicating $D_{\langle VT \rangle_{\rm sen}}$ (left-panel) or $R_{90\%}$ (right-panel). Spinning injection sets are labelled and shown as displaced circles. The blue and light blue shaded regions mark the template space encompassed by the GstLAL ($M<400~{\rm M}_\odot$) and PyCBC ($M<500~{M}_\odot$) template banks respectively.}				\label{fig:sensitive_distance_cwb_O1-O2}
\end{figure*}

	The left panel of Figure \ref{fig:sensitive_distance_cwb_O1-O2} shows the sensitive distance reach of our combined search toward our {17} targeted IMBHB sources represented in the $m_1-m_2$ plane (see Table~\ref{tab:tab23} in Appendix~\ref{sec:appeB} for a more detailed description). 
	We find an across-the-board improvement in the sensitive distance of our combined search compared to the 12 targeted sources reported in Ref.~\cite{Abbott:2017iws}.
	In particular, we find that the combined search is most sensitive to the $(100+100)~{\rm M}_\odot$ aligned-spin source, which can be observed up to ${1.8~{\rm Gpc}}$ and is an increase of more than {10\%} compared to the $1.6$~Gpc obtained in Ref.~\cite{Abbott:2017iws}.
	These improvements are the result of better detector sensitivity, the inclusion of higher modes in our injections, and significant improvements to the cWB search algorithm.
	As a general trend, our reach decreases for increasing mass ratio and for increasing total mass once this surpasses $\sim 200M_\odot$ .
	There are several reasons for this behaviour.
	First, the intrinsic amplitude of IMBHB signal decreases as the mass ratio decreases for a fixed total mass.
	Second, sources with small $q$ have a significant fraction of their power contained in their higher modes.
	Consequently, they are not well matched by our search templates, which only include the dominant quadrupole mode.
	Last, although the intrinsic luminosity of IMBHBs rises with total mass, the merger frequency decreases, and so signals persist in the detector sensitive frequency band for a very short duration.
	{This makes it difficult to distinguish them from noise transients. 
	This effect is evident from the roughly equivalent sensitive distances obtained for the $(60+60)~{\rm M}_\odot$ and $(100+100)~{\rm M}_\odot$ sources despite the significantly larger total mass of the latter.}

	 The right panel of Figure \ref{fig:sensitive_distance_cwb_O1-O2} shows the upper limits on the merger rate density of our 17 targeted IMBHB sources, which improve on those reported after O1 in Ref.~\cite{Abbott:2017iws}. 
	We set our most stringent upper limit at {$0.20$~${\rm Gpc}^{-3}~{\rm yr}^{-1}$} for equal-mass spin-aligned IMBHBs with component masses of $100~{\rm M}_\odot$ and aligned dimensionless spins of 0.8. 
	 By assuming a redshift-independent globular cluster (GC) density of $3~$GC~Mpc$^{-3}$~\cite{PortegiesZwart:2000}, we find that this upper limit is equivalent to {0.07}~GC$^{-1}$~Gyr$^{-1}$, an improvement of a factor of {$\sim 5$} over the $0.31$~GC$^{-1}$~Gyr$^{-1}$ that was reported in Ref.~\cite{Abbott:2017iws}.  
	We also observe that for all equal mass ratio IMBHB sources, the merger rate density upper limits are also impoved.
	The sources with unequal masses show larger improvement in the merger rate density as compared to previous result. 
	\section{Conclusions} 
	\label{sec:conclusion}
	
	We have conducted a search for IMBHBs in the data collected in the two observing runs of the Advanced LIGO and Virgo detectors.
	This search combined three analysis pipelines: two matched-filter algorithms GstLAL and PyCBC and
	the model-independent algorithm cWB.
	The PyCBC and cWB analyses use data from the Advanced LIGO detectors, and GstLAL uses data from the Advanced LIGO and Advanced Virgo detectors. 
	No IMBHB {detections were made} in this search.
	The loudest {candidate} event was found with a marginal $p$-value $\bar{P}=0.36$ in our combined search.
	A {detailed} detector characterization study of this event
	suggested that it is likely explained by the detector noise. 
	
	Given the null detection, we place upper limits on the merger rate density for 17 IMBHB systems.
	For estimation of the rate upper limits, we use NR waveforms provided by the SXS, RIT, and Georgia Tech groups that include higher modes in the gravitational-wave emission. 
	The reported rate limits are significantly more stringent than
	the previous result reported in Ref.~\cite{Abbott:2017iws}.
	In particular, the most stringent rate limit of {$0.20$ ${\rm Gpc}^{-3} {\rm yr}^{-1}$} placed on (100+100)~${\rm M}_\odot$ aligned spin IMBHB systems is an improvement of a factor of {$\sim 5$}.
	 This improvement is due to the combination of three factors: the increased sensitivity of our detector network, the improvements in the cWB search algorithm, and the incorporation of higher modes into our models for IMBHB signals.
	
	Anticipated increases of the network sensitivity in future runs,
	particularly at low frequency, and further improvement of the search 
	algorithms will place more stringent upper limits on the merger rate density
	of IMBHBs and may even result in the first definitive detection of an IMBH.

	\acknowledgments
	{The authors gratefully acknowledge the support of the United States
National Science Foundation (NSF) for the construction and operation of the
LIGO Laboratory and Advanced LIGO as well as the Science and Technology Facilities Council (STFC) of the
United Kingdom, the Max-Planck-Society (MPS), and the State of
Niedersachsen/Germany for support of the construction of Advanced LIGO 
and construction and operation of the GEO600 detector. 
Additional support for Advanced LIGO was provided by the Australian Research Council.
The authors gratefully acknowledge the Italian Istituto Nazionale di Fisica Nucleare (INFN),  
the French Centre National de la Recherche Scientifique (CNRS) and
the Foundation for Fundamental Research on Matter supported by the Netherlands Organisation for Scientific Research, 
for the construction and operation of the Virgo detector
and the creation and support  of the EGO consortium. 
The authors also gratefully acknowledge research support from these agencies as well as by 
the Council of Scientific and Industrial Research of India, 
the Department of Science and Technology, India,
the Science \& Engineering Research Board (SERB), India,
the Ministry of Human Resource Development, India,
the Spanish  Agencia Estatal de Investigaci\'on,
the Vicepresid\`encia i Conselleria d'Innovaci\'o, Recerca i Turisme and the Conselleria d'Educaci\'o i Universitat del Govern de les Illes Balears,
the Conselleria d'Educaci\'o, Investigaci\'o, Cultura i Esport de la Generalitat Valenciana,
the National Science Centre of Poland,
the Swiss National Science Foundation (SNSF),
the Russian Foundation for Basic Research, 
the Russian Science Foundation,
the European Commission,
the European Regional Development Funds (ERDF),
the Royal Society, 
the Scottish Funding Council, 
the Scottish Universities Physics Alliance, 
the Hungarian Scientific Research Fund (OTKA),
the Lyon Institute of Origins (LIO),
the Paris \^{I}le-de-France Region, 
the National Research, Development and Innovation Office Hungary (NKFIH), 
the National Research Foundation of Korea,
Industry Canada and the Province of Ontario through the Ministry of Economic Development and Innovation, 
the Natural Science and Engineering Research Council Canada,
the Canadian Institute for Advanced Research,
the Brazilian Ministry of Science, Technology, Innovations, and Communications,
the International Center for Theoretical Physics South American Institute for Fundamental Research (ICTP-SAIFR), 
the Research Grants Council of Hong Kong,
the National Natural Science Foundation of China (NSFC),
the Leverhulme Trust, 
the Research Corporation, 
the Ministry of Science and Technology (MOST), Taiwan
and
the Kavli Foundation.
The authors gratefully acknowledge the support of the NSF, STFC, MPS, INFN, CNRS and the
State of Niedersachsen/Germany for provision of computational resources.}

\appendix
	
	\section{Sensitive Distance reach and merger rate}
	\label{sec:sensivetive_distance_reach}
	In this appendix we provide further details on our method to compute the averaged spacetime volume observed by a search and its corresponding sensitive distance at a given significance threshold. In general, the averaged spacetime volume to which our searches are sensitive  is given by \cite{Abbott:2016drs,Abbott:2016nhf}:
	\begin{equation} 
	\langle VT\rangle_{\rm sen}  \; =T_0 \int dz  \; \; d \boldsymbol{\theta}  \; \; \frac{dV_{\rm c}}{dz} \; \; \frac{1}{1+z}  \; \; s(\boldsymbol{\theta})  \; \; f(z,\boldsymbol{\theta}).
	\label{eq:sens_general}
	\end{equation}	
	Here, $T_0$ is the length of the observation in the detector frame, and $V_{\rm c}(z)$ is the co-moving volume spanned by a sphere of redshift $z$.
	The function  $s({\bf \boldsymbol{\theta}})$ is the distribution of binary parameters ${\bf \boldsymbol{\theta}}$, and $0 \; \leq \; f(z,\boldsymbol{\theta}) \; \leq \; 1 $, where $f(z,\boldsymbol{\theta})$ denotes the fraction of injections with parameters $\boldsymbol{\theta}$ detected at a redshift $z$.

	In this {determination of sensitivity} we   
	 have two main limitations.
	First, the true population of IMBHBs in the Universe is unknown, so it prevents us from choosing a particular function $s({\bf \boldsymbol{\theta}})$.
	Second, numerical relativity waveforms cover a discrete parameter space in ${\bf\boldsymbol{\theta}}$. 
	For this reason, our study is focused on probing a discrete set of IMBHB classes with parameters $\{\theta_i\}$, described in Table~\ref{tab:tab23}. 
	Then the averaged space-time volume sensitivity of Eq.~\ref{eq:sens_general} can be approximated using a Monte-Carlo technique via
	\begin{equation}
	\label{svt}
	\langle VT  \rangle_{\rm sen} \sim \frac{N_{\rm rec}}{N_{\rm tot}}\langle V T\rangle_{\rm tot} \, .
	\end{equation}
	Here, $N_{\rm tot}$ is the total number of injections in a given set, which are distributed in redshift and source orientations as indicated in Sec. IIIA.
	$\langle VT\rangle_{\rm tot}$ is the total spacetime volume into which injections were distributed.
	$N_{\rm rec}$ is the number of recovered injections by the search, i.e. the number of injections assigned a value $\bar{P} \leq \bar{P}_0$, where $\bar{P}_0$ is in general some arbitrary threshold.
	In our case, we set $\bar{P}_0=0.36$, which is the $\bar{P}$ of our most significant event in our combined search.

	The corresponding sensitive distance reach is computed as
	\begin{equation}\label{sensitive_distance_reach}
	D_{\langle VT \rangle_{\rm sen}}= \left(\frac{3 \; \langle VT\rangle_{\rm sen}}{ 4\pi \; T_{\rm a}}\right)^{1/3},
	\end{equation}
		where $T_{\rm a}$ is the amount of time analyzed by the search.
		We estimated the $90~\%$ confidence upper limit in the merger rate density for selected simulated signal classes as given by 
		
			\begin{equation}\label{eq:rate}
			R_{90\%} = - \frac{ \ln(0.1)  }{\langle VT\rangle_{\rm sen}},
			\end{equation}
	where $\langle VT\rangle_{\rm sen}$ is estimated using the loudest event method and Equation \ref{svt}.

\section{Determining the p-value of the combined search}
\label{sec:pval}

In general, the $p$-value of the triggers of our combined search is given by 
\begin{equation}
	\label{pbar_m}
	\bar{P}=1-(1-P_{\rm min})^m, \
	\end{equation}
where $P_{\rm min}$	 is the minimum $p$-value reported by any of our three searches, and $m\in[1,3]$ is the trials factor.
The trials factor is $m=1$ if the three searches are fully correlated (for instance, if they are the same search) and three if they are fully independent.
In this work we adopt a conservative approach and choose $m=3$, omitting possible correlations between the three analysis pipelines.
Indeed, excluding the eleven detected GWs mentioned in Sec.~\ref{sec:SecIV}, none of the 123 events with ${\rm FAR}<100/{\rm yr}$ {was} common across the three pipelines.

	We note that while the significance of individual triggers depends on our particular choice for the trials factor $m$ applied in Eq.~\ref{pbar_m} (which we set to $m=3$), $N_{\rm rec}$ is independent of this choice.
	This is because every GW candidate output by the three analyses, including our loudest event, will have the same trials factor applied when combined into a single list, so that their relative ranking will remain unchanged (see Sec.~\ref{sec:combined}).
	Therefore, the numerical value of our upper limit is unaffected by our conservative choice of $m=3$, since any choice would yield the same $N_{\rm rec}$ and $\langle VT  \rangle_{\rm sen}$.
	
	As pointed out, since our choice of the trials factor affects the significance of individual triggers, our conservative approach may overly diminish the significance of prospective louder IMBHB triggers, and it may become important to make more accurate estimates of $m$ in the future.
	Since the lowest $p$-value reported by any of our individual analyses was $P = 0.14$, we conclude that our choice of $m$ does not impact our conclusion that no IMBHBs have been observed.

\section{Sensitive distance reach for individual search algorithms }
\label{sec:appeB}
In this appendix, we report and compare the sensitive distance reach of the three individual searches and the combined search at their respective loudest event thresholds (see Table \ref{tab:tab23}).
For the case of the individual searches, this threshold is set to $\bar{P}_{0}=0.14$, equal to the loudest (most significant) event found by cWB; for the combined search, this is set to $\bar{P}_{0}=0.36$. 
We control for differences in the amount of analyzed time by only considering common observing times in Table \ref{tab:tab23}, which allows for a more direct comparison between the searches.
 
	Table~\ref{tab:tab23} shows that cWB reports the largest sensitive distance reach to every IMBHB source considered.
	This is expected for sources with $M_{\rm tot}>500$~M$_\odot$, since the GstLAL and PyCBC template banks are bounded by a total mass 400~M$_\odot$ and 500~M$_\odot$, respectively.
	Additionally, since cWB is not limited by constraints on waveform morphology, it significantly outperforms matched-filter analyses in the large mass and small mass ratio regions of the parameter space that are covered by our analyses' template banks.
	This finding is consistent with Ref.~\cite{CalderonBustillo:2017skv}, since in that region of parameter space, signals are shorter and higher modes are more important.
	Ref.~\cite{CalderonBustillo:2017skv} also found that matched-filter searches outperform cWB in the low mass end of our parameter space. 
	Since then, however, cWB has undergone major improvements that have led to a sensitivity comparable to that of matched filter searches even for the lightest equal-mass systems considered in this analysis.
	
	GstLAL reports sensitive distance reaches that are lower than those found in Ref.~\cite{Abbott:2017iws}.
	This is the result of using a large bank {here} that was not specifically tuned and targeted for IMBHBs.
	Future searches will benefit from investigations into optimal template placement and binning as well as a return to a dedicated IMBH bank.
	
\begin{widetext}
	
	\begin{table}
		
		\begin{center}
			\begin{center}
				\setlength\extrarowheight{5pt}
				\begin{tabular}{ |c  | c | c | c | c |c | c | c |c |  }
					\hline 
					\:\: $m_1$ \: \: &\:\:  $m_2$  \:\:  &\:\:  spin\:\:  &\:\:  $M$ \:\:  & \qquad \qquad \qquad \qquad NR-simulation \qquad \qquad \qquad \qquad &  \multicolumn{4}{c|}{{$D_{\langle VT \rangle_{\rm sen}}$ (${\rm Gpc}$)}}   \\ 
					${\rm M}_\odot$ &  ${\rm M}_\odot$ & $\chi_{1,2}$ &  ${\rm M}_\odot$  & & cWB & GstLAL & PyCBC & combined \\ \hline
					
					60 & 60 & 0 &120 & SXS:BBH:0180, RIT:BBH:0198:n140, GT:0905 & 1.2& 1.2 & 1.2& 1.3 \\ 
					
					60 & 60 & 0.8 & 120 & SXS:BBH:0230, RIT:BBH:0063:n100, GT:0424  & 1.6 & 1.0 &1.5 &1.6 \\ 
					
					100 & 20 & 0 & 120 & SXS:BBH:0056, RIT:BBH0120:n140, GT:0906  & 0.72 & 0.69 & 0.70&0.76\\
					
					100 & 50 & 0 & 150 & SXS:BBH:0169, RIT:BBH:0117:n140, GT:0446  &1.2 & 0.79 & 1.1 & 1.2 \\
					
					100 & 100 & -0.8 & 200 & SXS:BBH:0154, RIT:BBH:0068:n100  & 1.1 & 1.0& 0.99 &  1.2 \\
					
					100 & 100 & 0 & 200 & SXS:BBH:0180, RIT:BBH:0198:n140,GT:0905  &1.4 & 0.90 & 1.3 &1.4\\
					
					100 & 100 & 0.8 & 200 & SXS:BBH:0230, RIT:BBH:0063:n100, GT:0424  &1.8 & 1.2 & 1.7 & 1.8  \\
					
					200 & 20 & 0 & 220 & RIT:BBH:Q10:n173, GT:0568  & 0.48& 0.30 &0.36 & 0.49\\ 
					
					200 & 50 & 0 & 250 & SXS:BBH:0182, RIT:BBH:0119:n140, GT:0454  & 0.85 &0.48 &0.67 & 0.87\\
					
					200 & 100 & 0 & 300 & SXS:BBH:0169, RIT:BBH:0117:n140, GT:0446  &1.1 & 0.59 &0.86 &1.1  \\
					
					300 & 50 & 0 & 350 & SXS:BBH:0181, RIT:BBH:0121:n140, GT:0604  &0.55 & 0.18 &0.27 & 0.56 \\
					
					200 & 200 & 0 & 400 & SXS:BBH:0180, RIT:BBH:0198:n140, GT:0905 &1.0 & 0.47 & 0.72&1.0 \\
					300 & 100 & 0 & 400 & SXS:BBH:0030, RIT:BBH:0102:n140, GT:0453  &0.78 & 0.23 & 0.34 & 0.78 \\
					
					400 & 40 & 0 & 440 & RIT:BBH:Q10:n173, GT:0568  &0.35 & 0.10 &0.16 & 0.35\\
					
					300 & 200 & 0 & 500 & RIT:BBH:0115:n140, GT:0477  & 0.79 & 0.16 & 0.14 & 0.79  \\
					
					300 & 300 & 0 & 600 & SXS:BBH:0180, RIT:BBH:0198:n140, GT:0905  & 0.61 & 0.09 & 0.18 & 0.61 \\
					
					400 & 400 & 0 & 800 & SXS:BBH:0180, RIT:BBH:0198:n140, GT:0905  &0.31 & 0.10 & 0.23 & 0.31\\\hline

				\end{tabular}
			\end{center}
			
		\end{center}
		
		\caption{\label{tab:tab23} The sensitive distance reach and the merger rate density calculated for the 17 targeted IMBHB sources considered in this study, whose intrinsic parameters are indicated in the first four columns. The fifth column indicates the numerical simulations used for each case, following the naming conventions of the corresponding NR groups. The next three columns report the sensitive distance reach for each of the individual analyses (cWB, GstLAL, and PyCBC), where we use the loudest event threshold of $P=0.14$ for each analysis for comparison purposes. The last column gives the sensitive distance reach from the combined search. To control for differences in the amount of analyzed time between individual analyses, we consider only common observed time across the three pipelines time, which yields $T_{\rm a}= 0.428$ years.}
	\end{table}
	
\end{widetext}

	\section{Loudest event parameter estimation}
	\label{sec:pe}

Despite the low significance of our loudest event, two characteristics motivated a detailed followup analysis.
On the one hand, initial parameter estimation put this trigger in the IMBHB region of the parameter space.
On the other, this trigger was observed by our matched filter analyses with an SNR of only $\sim6$, much lower than that recovered by cWB.
If this were a real GW, {this difference might} be indicative that the signal contained physics that our search templates omit (such as precession and higher modes), which would lead to a reduction of its SNR and  significance. 

To explore this {possibility}, we ran standard parameter estimation on this event using the same approximants used in Ref.~\cite{O1-O2-catalog-paper}, namely SEOBNRv4~\cite{Bohe:2016gbl} and IMRPhenomPv2 \cite{Hannam:2013oca}. Note that the latter approximant includes the effects of precession that our search templates omit.
For the precessing IMRPhenomP run, we assumed a spin magnitude prior uniform between 0 and 0.99, and spin orientations were isotropically distributed on the sphere;
for the spin-aligned SEOBNR waveforms, we used a spin prior such that the components of the spin aligned with the orbital angular momentum matched the prior used for the IMRPhenomP analysis.
Remarkably, the two analyses not only report broadly consistent parameter posterior distributions but they also report consistent SNRs of  $\sim 6$, in agreement with that reported by our matched filter searches.
The latter indicates that the low SNR obtained by our matched filter searches is not likely due to lack of precession in our templates.
Assuming this event is a compact binary, we recover a source-frame chirp mass of $70^{+24}_{-20}$ M$_{\odot}$, a source-frame total mass of $171^{+68}_{-48}$ $\rm{M}_{\odot}$, an effective inspiral spin of $0.19^{+0.44}_{-0.46}$, and a luminosity distance of $7.0^{+8.0}_{-4.2}$ {\rm Gpc}.
We also note that, given the lack of information about the spins, spin results are sensitive to the choice of prior.
Further parameter estimation was performed using the new SEOBNRv4HM \cite{Cotesta:2018fcv} approximant, which includes the impact of higher order modes. {The resulting} consistent parameter posteriors and no increase of the SNR, suggesting that the low SNR obtained by our matched filter searches is not likely due to lack of higher modes in our templates either.

We further conducted parameter estimation of this trigger by directly using numerical relativity waveforms of generic spin configurations and higher-modes with the RIFT algorithm~\cite{Lange:2017wki,2018arXiv180510457L}, which reported results consistent with those obtained by our waveform approximants.
In addition, the event was also reconstructed using the model agnostic algorithm BayesWave~\cite{Cornish:2014kda,bayeswave}, which reported an SNR consistent with those obtained by our templates.

In summary, detailed followup of this event suggests that, in the most optimistic scenario, this trigger would be the combination of a weak IMBHB signal plus a noise transient with power detected by cWB (see Sec.~\ref{sec:SecIV}), raising the significance of the underlying IMBHB signal.
Since the resulting event has a marginal significance, the underlying IMBHB trigger would be even less significant.
Hence, we conclude that this event is best explained by detector noise.

\bibliography{imbhb_O1-O2.bib}

\begin{thebibliography}{103}%
\makeatletter
\providecommand \@ifxundefined [1]{%
 \@ifx{#1\undefined}
}%
\providecommand \@ifnum [1]{%
 \ifnum #1\expandafter \@firstoftwo
 \else \expandafter \@secondoftwo
 \fi
}%
\providecommand \@ifx [1]{%
 \ifx #1\expandafter \@firstoftwo
 \else \expandafter \@secondoftwo
 \fi
}%
\providecommand \natexlab [1]{#1}%
\providecommand \enquote  [1]{``#1''}%
\providecommand \bibnamefont  [1]{#1}%
\providecommand \bibfnamefont [1]{#1}%
\providecommand \citenamefont [1]{#1}%
\providecommand \href@noop [0]{\@secondoftwo}%
\providecommand \href [0]{\begingroup \@sanitize@url \@href}%
\providecommand \@href[1]{\@@startlink{#1}\@@href}%
\providecommand \@@href[1]{\endgroup#1\@@endlink}%
\providecommand \@sanitize@url [0]{\catcode `\\12\catcode `\$12\catcode
  `\&12\catcode `\#12\catcode `\^12\catcode `\_12\catcode `\%12\relax}%
\providecommand \@@startlink[1]{}%
\providecommand \@@endlink[0]{}%
\providecommand \url  [0]{\begingroup\@sanitize@url \@url }%
\providecommand \@url [1]{\endgroup\@href {#1}{\urlprefix }}%
\providecommand \urlprefix  [0]{URL }%
\providecommand \Eprint [0]{\href }%
\providecommand \doibase [0]{http://dx.doi.org/}%
\providecommand \selectlanguage [0]{\@gobble}%
\providecommand \bibinfo  [0]{\@secondoftwo}%
\providecommand \bibfield  [0]{\@secondoftwo}%
\providecommand \translation [1]{[#1]}%
\providecommand \BibitemOpen [0]{}%
\providecommand \bibitemStop [0]{}%
\providecommand \bibitemNoStop [0]{.\EOS\space}%
\providecommand \EOS [0]{\spacefactor3000\relax}%
\providecommand \BibitemShut  [1]{\csname bibitem#1\endcsname}%
\let\auto@bib@innerbib\@empty
\bibitem [{\citenamefont {Abbott}\ \emph
  {et~al.}(2016{\natexlab{a}})\citenamefont {Abbott} \emph
  {et~al.}}]{GW150914-DETECTION}%
  \BibitemOpen
  \bibfield  {author} {\bibinfo {author} {\bibfnamefont {B.~P.}\ \bibnamefont
  {Abbott}} \emph {et~al.} (\bibinfo {collaboration} {LIGO Scientific
  Collaboration, Virgo Collaboration}),\ }\href {\doibase
  10.1103/PhysRevLett.116.061102} {\bibfield  {journal} {\bibinfo  {journal}
  {Phys. Rev. Lett.}\ }\textbf {\bibinfo {volume} {116}},\ \bibinfo {pages}
  {061102} (\bibinfo {year} {2016}{\natexlab{a}})},\ \Eprint
  {http://arxiv.org/abs/1602.03837} {arXiv:1602.03837 [gr-qc]} \BibitemShut
  {NoStop}%
\bibitem [{\citenamefont {Abbott}\ \emph
  {et~al.}(2016{\natexlab{b}})\citenamefont {Abbott} \emph
  {et~al.}}]{GW151226-DETECTION}%
  \BibitemOpen
  \bibfield  {author} {\bibinfo {author} {\bibfnamefont {B.~P.}\ \bibnamefont
  {Abbott}} \emph {et~al.} (\bibinfo {collaboration} {LIGO Scientific
  Collaboration, Virgo Collaboration}),\ }\href {\doibase
  10.1103/PhysRevLett.116.241103} {\bibfield  {journal} {\bibinfo  {journal}
  {Phys. Rev. Lett.}\ }\textbf {\bibinfo {volume} {116}},\ \bibinfo {pages}
  {241103} (\bibinfo {year} {2016}{\natexlab{b}})},\ \Eprint
  {http://arxiv.org/abs/1606.04855} {arXiv:1606.04855 [gr-qc]} \BibitemShut
  {NoStop}%
\bibitem [{\citenamefont {Abbott}\ \emph
  {et~al.}(2017{\natexlab{a}})\citenamefont {Abbott} \emph
  {et~al.}}]{GW170104-DETECTION}%
  \BibitemOpen
  \bibfield  {author} {\bibinfo {author} {\bibfnamefont {B.~P.}\ \bibnamefont
  {Abbott}} \emph {et~al.} (\bibinfo {collaboration} {LIGO Scientific
  Collaboration, Virgo Collaboration}),\ }\href {\doibase
  10.1103/PhysRevLett.118.221101} {\bibfield  {journal} {\bibinfo  {journal}
  {Phys. Rev. Lett.}\ }\textbf {\bibinfo {volume} {118}},\ \bibinfo {pages}
  {221101} (\bibinfo {year} {2017}{\natexlab{a}})},\ \Eprint
  {http://arxiv.org/abs/1706.01812} {arXiv:1706.01812 [gr-qc]} \BibitemShut
  {NoStop}%
\bibitem [{\citenamefont {Abbott}\ \emph
  {et~al.}(2017{\natexlab{b}})\citenamefont {Abbott} \emph
  {et~al.}}]{GW170608-DETECTION}%
  \BibitemOpen
  \bibfield  {author} {\bibinfo {author} {\bibfnamefont {B.~P.}\ \bibnamefont
  {Abbott}} \emph {et~al.} (\bibinfo {collaboration} {LIGO Scientific
  Collaboration, Virgo Collaboration}),\ }\href {\doibase
  10.3847/2041-8213/aa9f0c} {\bibfield  {journal} {\bibinfo  {journal}
  {Astrophys. J.}\ }\textbf {\bibinfo {volume} {851}},\ \bibinfo {pages} {L35}
  (\bibinfo {year} {2017}{\natexlab{b}})},\ \Eprint
  {http://arxiv.org/abs/1711.05578} {arXiv:1711.05578 [astro-ph.HE]}
  \BibitemShut {NoStop}%
\bibitem [{\citenamefont {Abbott}\ \emph
  {et~al.}(2017{\natexlab{c}})\citenamefont {Abbott} \emph
  {et~al.}}]{GW170817-DETECTION}%
  \BibitemOpen
  \bibfield  {author} {\bibinfo {author} {\bibfnamefont {B.}~\bibnamefont
  {Abbott}} \emph {et~al.} (\bibinfo {collaboration} {LIGO Scientific
  Collaboration, Virgo Collaboration}),\ }\href {\doibase
  10.1103/PhysRevLett.119.161101} {\bibfield  {journal} {\bibinfo  {journal}
  {Phys. Rev. Lett.}\ }\textbf {\bibinfo {volume} {119}},\ \bibinfo {pages}
  {161101} (\bibinfo {year} {2017}{\natexlab{c}})},\ \Eprint
  {http://arxiv.org/abs/1710.05832} {arXiv:1710.05832 [gr-qc]} \BibitemShut
  {NoStop}%
\bibitem [{\citenamefont {Abbott}\ \emph
  {et~al.}(2017{\natexlab{d}})\citenamefont {Abbott} \emph
  {et~al.}}]{GW170814-DETECTION}%
  \BibitemOpen
  \bibfield  {author} {\bibinfo {author} {\bibfnamefont {B.~P.}\ \bibnamefont
  {Abbott}} \emph {et~al.} (\bibinfo {collaboration} {LIGO Scientific
  Collaboration, Virgo Collaboration}),\ }\href {\doibase
  10.1103/PhysRevLett.119.141101} {\bibfield  {journal} {\bibinfo  {journal}
  {Phys. Rev. Lett.}\ }\textbf {\bibinfo {volume} {119}},\ \bibinfo {pages}
  {141101} (\bibinfo {year} {2017}{\natexlab{d}})},\ \Eprint
  {http://arxiv.org/abs/1709.09660} {arXiv:1709.09660 [gr-qc]} \BibitemShut
  {NoStop}%
\bibitem [{\citenamefont {Abbott}\ \emph
  {et~al.}(2016{\natexlab{c}})\citenamefont {Abbott} \emph
  {et~al.}}]{TheLIGOScientific:2016pea}%
  \BibitemOpen
  \bibfield  {author} {\bibinfo {author} {\bibfnamefont {B.~P.}\ \bibnamefont
  {Abbott}} \emph {et~al.} (\bibinfo {collaboration} {LIGO Scientific
  Collaboration, Virgo Collaboration}),\ }\href {\doibase
  10.1103/PhysRevX.6.041015, 10.1103/PhysRevX.8.039903} {\bibfield  {journal}
  {\bibinfo  {journal} {Phys. Rev.}\ }\textbf {\bibinfo {volume} {X6}},\
  \bibinfo {pages} {041015} (\bibinfo {year} {2016}{\natexlab{c}})},\ \bibinfo
  {note} {[erratum: Phys. Rev.X8,no.3,039903(2018)]},\ \Eprint
  {http://arxiv.org/abs/1606.04856} {arXiv:1606.04856 [gr-qc]} \BibitemShut
  {NoStop}%
\bibitem [{\citenamefont {Abbott}\ \emph
  {et~al.}(2018{\natexlab{a}})\citenamefont {Abbott} \emph
  {et~al.}}]{O1-O2-catalog-paper}%
  \BibitemOpen
  \bibfield  {author} {\bibinfo {author} {\bibfnamefont {B.~P.}\ \bibnamefont
  {Abbott}} \emph {et~al.} (\bibinfo {collaboration} {LIGO Scientific
  Collaboration and Virgo Collaboration}),\ }\href@noop {} {\  (\bibinfo {year}
  {2018}{\natexlab{a}})},\ \Eprint {http://arxiv.org/abs/1811.12907}
  {arXiv:1811.12907 [astro-ph.HE]} \BibitemShut {NoStop}%
\bibitem [{\citenamefont {Abbott}\ \emph
  {et~al.}(2018{\natexlab{b}})\citenamefont {Abbott} \emph
  {et~al.}}]{O1-O2-rate_paper}%
  \BibitemOpen
  \bibfield  {author} {\bibinfo {author} {\bibfnamefont {B.~P.}\ \bibnamefont
  {Abbott}} \emph {et~al.} (\bibinfo {collaboration} {LIGO Scientific
  Collaboration and Virgo Collaboration}),\ }\href@noop {} {\  (\bibinfo {year}
  {2018}{\natexlab{b}})},\ \Eprint {http://arxiv.org/abs/1811.12940}
  {arXiv:1811.12940 [astro-ph.HE]} \BibitemShut {NoStop}%
\bibitem [{\citenamefont {Spera}\ and\ \citenamefont
  {Mapelli}(2017)}]{Spera:2017fyx}%
  \BibitemOpen
  \bibfield  {author} {\bibinfo {author} {\bibfnamefont {M.}~\bibnamefont
  {Spera}}\ and\ \bibinfo {author} {\bibfnamefont {M.}~\bibnamefont
  {Mapelli}},\ }\href {\doibase 10.1093/mnras/stx1576} {\bibfield  {journal}
  {\bibinfo  {journal} {Mon. Not. Roy. Astron. Soc.}\ }\textbf {\bibinfo
  {volume} {470}},\ \bibinfo {pages} {4739} (\bibinfo {year} {2017})},\ \Eprint
  {http://arxiv.org/abs/1706.06109} {arXiv:1706.06109 [astro-ph.SR]}
  \BibitemShut {NoStop}%
\bibitem [{\citenamefont {Woosley}(2017)}]{Woosley:2016hmi}%
  \BibitemOpen
  \bibfield  {author} {\bibinfo {author} {\bibfnamefont {S.~E.}\ \bibnamefont
  {Woosley}},\ }\href {\doibase 10.3847/1538-4357/836/2/244} {\bibfield
  {journal} {\bibinfo  {journal} {Astrophys. J.}\ }\textbf {\bibinfo {volume}
  {836}},\ \bibinfo {pages} {244} (\bibinfo {year} {2017})},\ \Eprint
  {http://arxiv.org/abs/1608.08939} {arXiv:1608.08939 [astro-ph.HE]}
  \BibitemShut {NoStop}%
\bibitem [{\citenamefont {Giacobbo}\ \emph {et~al.}(2018)\citenamefont
  {Giacobbo}, \citenamefont {Mapelli},\ and\ \citenamefont
  {Spera}}]{Giacobbo:2017qhh}%
  \BibitemOpen
  \bibfield  {author} {\bibinfo {author} {\bibfnamefont {N.}~\bibnamefont
  {Giacobbo}}, \bibinfo {author} {\bibfnamefont {M.}~\bibnamefont {Mapelli}}, \
  and\ \bibinfo {author} {\bibfnamefont {M.}~\bibnamefont {Spera}},\ }\href
  {\doibase 10.1093/mnras/stx2933} {\bibfield  {journal} {\bibinfo  {journal}
  {Mon. Not. Roy. Astron. Soc.}\ }\textbf {\bibinfo {volume} {474}},\ \bibinfo
  {pages} {2959} (\bibinfo {year} {2018})},\ \Eprint
  {http://arxiv.org/abs/1711.03556} {arXiv:1711.03556 [astro-ph.SR]}
  \BibitemShut {NoStop}%
\bibitem [{\citenamefont {Belczynski}\ \emph {et~al.}(2016)\citenamefont
  {Belczynski} \emph {et~al.}}]{Belczynski:2016jno}%
  \BibitemOpen
  \bibfield  {author} {\bibinfo {author} {\bibfnamefont {K.}~\bibnamefont
  {Belczynski}} \emph {et~al.},\ }\href {\doibase 10.1051/0004-6361/201628980}
  {\bibfield  {journal} {\bibinfo  {journal} {Astron. Astrophys.}\ }\textbf
  {\bibinfo {volume} {594}},\ \bibinfo {pages} {A97} (\bibinfo {year}
  {2016})},\ \Eprint {http://arxiv.org/abs/1607.03116} {arXiv:1607.03116
  [astro-ph.HE]} \BibitemShut {NoStop}%
\bibitem [{\citenamefont {{Mezcua}}(2017)}]{Merzcua:2017}%
  \BibitemOpen
  \bibfield  {author} {\bibinfo {author} {\bibfnamefont {M.}~\bibnamefont
  {{Mezcua}}},\ }\href {\doibase 10.1142/S021827181730021X} {\bibfield
  {journal} {\bibinfo  {journal} {International Journal of Modern Physics D}\
  }\textbf {\bibinfo {volume} {26}},\ \bibinfo {eid} {1730021} (\bibinfo {year}
  {2017})},\ \Eprint {http://arxiv.org/abs/1705.09667} {arXiv:1705.09667}
  \BibitemShut {NoStop}%
\bibitem [{\citenamefont {{Koliopanos}}(2017)}]{2017mbhe.confE..51K}%
  \BibitemOpen
  \bibfield  {author} {\bibinfo {author} {\bibfnamefont {F.}~\bibnamefont
  {{Koliopanos}}},\ }in\ \href@noop {} {\emph {\bibinfo {booktitle}
  {Proceedings of the XII Multifrequency Behaviour of High Energy Cosmic
  Sources Workshop. 12-17 June}}}\ (\bibinfo {year} {2017})\ p.~\bibinfo
  {pages} {51},\ \Eprint {http://arxiv.org/abs/1801.01095} {arXiv:1801.01095
  [astro-ph.GA]} \BibitemShut {NoStop}%
\bibitem [{\citenamefont {Kiziltan}\ \emph {et~al.}()\citenamefont {Kiziltan},
  \citenamefont {Baumgardt},\ and\ \citenamefont {{Loeb}}}]{Kiziltan:imbh}%
  \BibitemOpen
  \bibfield  {author} {\bibinfo {author} {\bibfnamefont {B.}~\bibnamefont
  {Kiziltan}}, \bibinfo {author} {\bibfnamefont {H.}~\bibnamefont {Baumgardt}},
  \ and\ \bibinfo {author} {\bibfnamefont {A.}~\bibnamefont {{Loeb}}},\
  }\href@noop {} {\enquote {\bibinfo {title} {An intermediate-mass black hole
  in the centre of the globular cluster 47 tucanae},}\ }\BibitemShut {NoStop}%
\bibitem [{\citenamefont {den Brok}\ \emph {et~al.}(2015)\citenamefont {den
  Brok} \emph {et~al.}}]{denBrok}%
  \BibitemOpen
  \bibfield  {author} {\bibinfo {author} {\bibnamefont {den Brok}} \emph
  {et~al.},\ }\href@noop {} {\bibfield  {journal} {\bibinfo  {journal} {The
  Astrophysical Journal}\ }\textbf {\bibinfo {volume} {809}} (\bibinfo {year}
  {2015})}\BibitemShut {NoStop}%
\bibitem [{\citenamefont {Atakan~Gurkan}\ \emph {et~al.}(2004)\citenamefont
  {Atakan~Gurkan}, \citenamefont {Freitag},\ and\ \citenamefont
  {Rasio}}]{AtakanGurkan:2003hm}%
  \BibitemOpen
  \bibfield  {author} {\bibinfo {author} {\bibfnamefont {M.}~\bibnamefont
  {Atakan~Gurkan}}, \bibinfo {author} {\bibfnamefont {M.}~\bibnamefont
  {Freitag}}, \ and\ \bibinfo {author} {\bibfnamefont {F.~A.}\ \bibnamefont
  {Rasio}},\ }\href {\doibase 10.1086/381968} {\bibfield  {journal} {\bibinfo
  {journal} {Astrophys. J.}\ }\textbf {\bibinfo {volume} {604}},\ \bibinfo
  {pages} {632} (\bibinfo {year} {2004})},\ \Eprint
  {http://arxiv.org/abs/astro-ph/0308449} {arXiv:astro-ph/0308449 [astro-ph]}
  \BibitemShut {NoStop}%
\bibitem [{\citenamefont {Kormendy}\ and\ \citenamefont
  {Ho}(2013)}]{Kormendy:2013dxa}%
  \BibitemOpen
  \bibfield  {author} {\bibinfo {author} {\bibfnamefont {J.}~\bibnamefont
  {Kormendy}}\ and\ \bibinfo {author} {\bibfnamefont {L.~C.}\ \bibnamefont
  {Ho}},\ }\href {\doibase 10.1146/annurev-astro-082708-101811} {\bibfield
  {journal} {\bibinfo  {journal} {Ann. Rev. Astron. Astrophys.}\ }\textbf
  {\bibinfo {volume} {51}},\ \bibinfo {pages} {511} (\bibinfo {year} {2013})},\
  \Eprint {http://arxiv.org/abs/1304.7762} {arXiv:1304.7762 [astro-ph.CO]}
  \BibitemShut {NoStop}%
\bibitem [{\citenamefont {Graham}(2012)}]{Graham:2012bs}%
  \BibitemOpen
  \bibfield  {author} {\bibinfo {author} {\bibfnamefont {A.~W.}\ \bibnamefont
  {Graham}},\ }\href {\doibase 10.1088/0004-637X/746/1/113} {\bibfield
  {journal} {\bibinfo  {journal} {Astrophys. J.}\ }\textbf {\bibinfo {volume}
  {746}},\ \bibinfo {pages} {113} (\bibinfo {year} {2012})},\ \Eprint
  {http://arxiv.org/abs/1202.1878} {arXiv:1202.1878 [astro-ph.CO]} \BibitemShut
  {NoStop}%
\bibitem [{\citenamefont {{Graham}}\ and\ \citenamefont
  {{Scott}}(2013)}]{2013ApJ...764..151G}%
  \BibitemOpen
  \bibfield  {author} {\bibinfo {author} {\bibfnamefont {A.~W.}\ \bibnamefont
  {{Graham}}}\ and\ \bibinfo {author} {\bibfnamefont {N.}~\bibnamefont
  {{Scott}}},\ }\href {\doibase 10.1088/0004-637X/764/2/151} {\bibfield
  {journal} {\bibinfo  {journal} {Astrophys. J.}\ }\textbf {\bibinfo {volume}
  {764}},\ \bibinfo {eid} {151} (\bibinfo {year} {2013})},\ \Eprint
  {http://arxiv.org/abs/1211.3199} {arXiv:1211.3199} \BibitemShut {NoStop}%
\bibitem [{\citenamefont {van~den Bosch}\ \emph {et~al.}(2006)\citenamefont
  {van~den Bosch}, \citenamefont {de~Zeeuw}, \citenamefont {Gebhardt},
  \citenamefont {Noyola},\ and\ \citenamefont {van~de
  Ven}}]{vandenBosch:2005uk}%
  \BibitemOpen
  \bibfield  {author} {\bibinfo {author} {\bibfnamefont {R.}~\bibnamefont
  {van~den Bosch}}, \bibinfo {author} {\bibfnamefont {T.}~\bibnamefont
  {de~Zeeuw}}, \bibinfo {author} {\bibfnamefont {K.}~\bibnamefont {Gebhardt}},
  \bibinfo {author} {\bibfnamefont {E.}~\bibnamefont {Noyola}}, \ and\ \bibinfo
  {author} {\bibfnamefont {G.}~\bibnamefont {van~de Ven}},\ }\href {\doibase
  10.1086/500644} {\bibfield  {journal} {\bibinfo  {journal} {Astrophys. J.}\
  }\textbf {\bibinfo {volume} {641}},\ \bibinfo {pages} {852} (\bibinfo {year}
  {2006})},\ \Eprint {http://arxiv.org/abs/astro-ph/0512503}
  {arXiv:astro-ph/0512503 [astro-ph]} \BibitemShut {NoStop}%
\bibitem [{\citenamefont {van~der Marel}\ and\ \citenamefont
  {Anderson}(2010)}]{vanderMarel:2009mc}%
  \BibitemOpen
  \bibfield  {author} {\bibinfo {author} {\bibfnamefont {R.~P.}\ \bibnamefont
  {van~der Marel}}\ and\ \bibinfo {author} {\bibfnamefont {J.}~\bibnamefont
  {Anderson}},\ }\href {\doibase 10.1088/0004-637X/710/2/1063} {\bibfield
  {journal} {\bibinfo  {journal} {Astrophys. J.}\ }\textbf {\bibinfo {volume}
  {710}},\ \bibinfo {pages} {1063} (\bibinfo {year} {2010})},\ \Eprint
  {http://arxiv.org/abs/0905.0638} {arXiv:0905.0638 [astro-ph.GA]} \BibitemShut
  {NoStop}%
\bibitem [{\citenamefont {Anderson}\ and\ \citenamefont {van~der
  Marel}(2010)}]{Anderson:2009ku}%
  \BibitemOpen
  \bibfield  {author} {\bibinfo {author} {\bibfnamefont {J.}~\bibnamefont
  {Anderson}}\ and\ \bibinfo {author} {\bibfnamefont {R.~P.}\ \bibnamefont
  {van~der Marel}},\ }\href {\doibase 10.1088/0004-637X/710/2/1032} {\bibfield
  {journal} {\bibinfo  {journal} {Astrophys. J.}\ }\textbf {\bibinfo {volume}
  {710}},\ \bibinfo {pages} {1032} (\bibinfo {year} {2010})},\ \Eprint
  {http://arxiv.org/abs/0905.0627} {arXiv:0905.0627 [astro-ph.GA]} \BibitemShut
  {NoStop}%
\bibitem [{\citenamefont {Baumgardt}()}]{Baumgart:2017imbh}%
  \BibitemOpen
  \bibfield  {author} {\bibinfo {author} {\bibfnamefont {H.}~\bibnamefont
  {Baumgardt}},\ }\href@noop {} {\enquote {\bibinfo {title} {N-body modeling of
  globular clusters: Masses, mass-to-light ratios and intermediate-mass black
  holes},}\ }\BibitemShut {NoStop}%
\bibitem [{\citenamefont {Baumgardt}\ \emph {et~al.}(2003)\citenamefont
  {Baumgardt}, \citenamefont {Hut}, \citenamefont {Makino}, \citenamefont
  {McMillan},\ and\ \citenamefont {Portegies~Zwart}}]{Baumgardt:2002rc}%
  \BibitemOpen
  \bibfield  {author} {\bibinfo {author} {\bibfnamefont {H.}~\bibnamefont
  {Baumgardt}}, \bibinfo {author} {\bibfnamefont {P.}~\bibnamefont {Hut}},
  \bibinfo {author} {\bibfnamefont {J.}~\bibnamefont {Makino}}, \bibinfo
  {author} {\bibfnamefont {S.}~\bibnamefont {McMillan}}, \ and\ \bibinfo
  {author} {\bibfnamefont {S.~F.}\ \bibnamefont {Portegies~Zwart}},\ }\href
  {\doibase 10.1086/367537} {\bibfield  {journal} {\bibinfo  {journal}
  {Astrophys. J.}\ }\textbf {\bibinfo {volume} {582}},\ \bibinfo {pages} {L21}
  (\bibinfo {year} {2003})},\ \Eprint {http://arxiv.org/abs/astro-ph/0210133}
  {arXiv:astro-ph/0210133 [astro-ph]} \BibitemShut {NoStop}%
\bibitem [{\citenamefont {Remillard}\ and\ \citenamefont
  {McClintock}(2006)}]{doi:10.1146/annurev.astro.44.051905.092532}%
  \BibitemOpen
  \bibfield  {author} {\bibinfo {author} {\bibfnamefont {R.~A.}\ \bibnamefont
  {Remillard}}\ and\ \bibinfo {author} {\bibfnamefont {J.~E.}\ \bibnamefont
  {McClintock}},\ }\href {\doibase 10.1146/annurev.astro.44.051905.092532}
  {\bibfield  {journal} {\bibinfo  {journal} {Annual Review of Astronomy and
  Astrophysics}\ }\textbf {\bibinfo {volume} {44}},\ \bibinfo {pages} {49}
  (\bibinfo {year} {2006})},\ \Eprint
  {http://arxiv.org/abs/https://doi.org/10.1146/annurev.astro.44.051905.092532}
  {https://doi.org/10.1146/annurev.astro.44.051905.092532} \BibitemShut
  {NoStop}%
\bibitem [{\citenamefont {Pasham}\ \emph {et~al.}(2015)\citenamefont {Pasham},
  \citenamefont {Strohmayer},\ and\ \citenamefont
  {Mushotzky}}]{Pasham:2015tca}%
  \BibitemOpen
  \bibfield  {author} {\bibinfo {author} {\bibfnamefont {D.~R.}\ \bibnamefont
  {Pasham}}, \bibinfo {author} {\bibfnamefont {T.~E.}\ \bibnamefont
  {Strohmayer}}, \ and\ \bibinfo {author} {\bibfnamefont {R.~F.}\ \bibnamefont
  {Mushotzky}},\ }\href {\doibase 10.1038/nature13710} {\bibfield  {journal}
  {\bibinfo  {journal} {Nature}\ }\textbf {\bibinfo {volume} {513}},\ \bibinfo
  {pages} {74} (\bibinfo {year} {2015})},\ \Eprint
  {http://arxiv.org/abs/1501.03180} {arXiv:1501.03180 [astro-ph.HE]}
  \BibitemShut {NoStop}%
\bibitem [{\citenamefont {Kaaret}\ \emph {et~al.}(2017)\citenamefont {Kaaret},
  \citenamefont {Feng},\ and\ \citenamefont {Roberts}}]{Kaaret:2017tcn}%
  \BibitemOpen
  \bibfield  {author} {\bibinfo {author} {\bibfnamefont {P.}~\bibnamefont
  {Kaaret}}, \bibinfo {author} {\bibfnamefont {H.}~\bibnamefont {Feng}}, \ and\
  \bibinfo {author} {\bibfnamefont {T.~P.}\ \bibnamefont {Roberts}},\ }\href
  {\doibase 10.1146/annurev-astro-091916-055259} {\bibfield  {journal}
  {\bibinfo  {journal} {Ann. Rev. Astron. Astrophys.}\ }\textbf {\bibinfo
  {volume} {55}},\ \bibinfo {pages} {303} (\bibinfo {year} {2017})},\ \Eprint
  {http://arxiv.org/abs/1703.10728} {arXiv:1703.10728 [astro-ph.HE]}
  \BibitemShut {NoStop}%
\bibitem [{\citenamefont {Farrell}\ \emph {et~al.}(2009)\citenamefont
  {Farrell}, \citenamefont {Webb}, \citenamefont {Barret}, \citenamefont
  {Godet},\ and\ \citenamefont {Rodrigues}}]{Farrell:2010bf}%
  \BibitemOpen
  \bibfield  {author} {\bibinfo {author} {\bibfnamefont {S.}~\bibnamefont
  {Farrell}}, \bibinfo {author} {\bibfnamefont {N.}~\bibnamefont {Webb}},
  \bibinfo {author} {\bibfnamefont {D.}~\bibnamefont {Barret}}, \bibinfo
  {author} {\bibfnamefont {O.}~\bibnamefont {Godet}}, \ and\ \bibinfo {author}
  {\bibfnamefont {J.}~\bibnamefont {Rodrigues}},\ }\href {\doibase
  10.1038/nature08083} {\bibfield  {journal} {\bibinfo  {journal} {Nature}\
  }\textbf {\bibinfo {volume} {460}},\ \bibinfo {pages} {73} (\bibinfo {year}
  {2009})},\ \Eprint {http://arxiv.org/abs/1001.0567} {arXiv:1001.0567
  [astro-ph.HE]} \BibitemShut {NoStop}%
\bibitem [{\citenamefont {Madau}\ and\ \citenamefont
  {Rees}(2001)}]{Madau:2001sc}%
  \BibitemOpen
  \bibfield  {author} {\bibinfo {author} {\bibfnamefont {P.}~\bibnamefont
  {Madau}}\ and\ \bibinfo {author} {\bibfnamefont {M.~J.}\ \bibnamefont
  {Rees}},\ }\href {\doibase 10.1086/319848} {\bibfield  {journal} {\bibinfo
  {journal} {Astrophys. J.}\ }\textbf {\bibinfo {volume} {551}},\ \bibinfo
  {pages} {L27} (\bibinfo {year} {2001})},\ \Eprint
  {http://arxiv.org/abs/astro-ph/0101223} {arXiv:astro-ph/0101223 [astro-ph]}
  \BibitemShut {NoStop}%
\bibitem [{\citenamefont {Schneider}\ \emph {et~al.}(2002)\citenamefont
  {Schneider}, \citenamefont {Ferrara}, \citenamefont {Natarajan},\ and\
  \citenamefont {Omukai}}]{Schneider:2001bu}%
  \BibitemOpen
  \bibfield  {author} {\bibinfo {author} {\bibfnamefont {R.}~\bibnamefont
  {Schneider}}, \bibinfo {author} {\bibfnamefont {A.}~\bibnamefont {Ferrara}},
  \bibinfo {author} {\bibfnamefont {P.}~\bibnamefont {Natarajan}}, \ and\
  \bibinfo {author} {\bibfnamefont {K.}~\bibnamefont {Omukai}},\ }\href
  {\doibase 10.1086/339917} {\bibfield  {journal} {\bibinfo  {journal}
  {Astrophys. J.}\ }\textbf {\bibinfo {volume} {571}},\ \bibinfo {pages} {30}
  (\bibinfo {year} {2002})},\ \Eprint {http://arxiv.org/abs/astro-ph/0111341}
  {arXiv:astro-ph/0111341 [astro-ph]} \BibitemShut {NoStop}%
\bibitem [{\citenamefont {Ryu}\ \emph {et~al.}(2016)\citenamefont {Ryu},
  \citenamefont {Tanaka}, \citenamefont {Perna},\ and\ \citenamefont
  {Haiman}}]{Ryu:2016dou}%
  \BibitemOpen
  \bibfield  {author} {\bibinfo {author} {\bibfnamefont {T.}~\bibnamefont
  {Ryu}}, \bibinfo {author} {\bibfnamefont {T.~L.}\ \bibnamefont {Tanaka}},
  \bibinfo {author} {\bibfnamefont {R.}~\bibnamefont {Perna}}, \ and\ \bibinfo
  {author} {\bibfnamefont {Z.}~\bibnamefont {Haiman}},\ }\href {\doibase
  10.1093/mnras/stw1241} {\bibfield  {journal} {\bibinfo  {journal} {Mon. Not.
  Roy. Astron. Soc.}\ }\textbf {\bibinfo {volume} {460}},\ \bibinfo {pages}
  {4122} (\bibinfo {year} {2016})},\ \Eprint {http://arxiv.org/abs/1603.08513}
  {arXiv:1603.08513 [astro-ph.GA]} \BibitemShut {NoStop}%
\bibitem [{\citenamefont {{Kinugawa}}\ \emph {et~al.}(2014)\citenamefont
  {{Kinugawa}}, \citenamefont {{Inayoshi}}, \citenamefont {{Hotokezaka}},
  \citenamefont {{Nakauchi}},\ and\ \citenamefont
  {{Nakamura}}}]{2014MNRAS.442.2963K}%
  \BibitemOpen
  \bibfield  {author} {\bibinfo {author} {\bibfnamefont {T.}~\bibnamefont
  {{Kinugawa}}}, \bibinfo {author} {\bibfnamefont {K.}~\bibnamefont
  {{Inayoshi}}}, \bibinfo {author} {\bibfnamefont {K.}~\bibnamefont
  {{Hotokezaka}}}, \bibinfo {author} {\bibfnamefont {D.}~\bibnamefont
  {{Nakauchi}}}, \ and\ \bibinfo {author} {\bibfnamefont {T.}~\bibnamefont
  {{Nakamura}}},\ }\href {\doibase 10.1093/mnras/stu1022} {\bibfield  {journal}
  {\bibinfo  {journal} {Mon. Not. Roy. Astron. Soc.}\ }\textbf {\bibinfo
  {volume} {442}},\ \bibinfo {pages} {2963} (\bibinfo {year} {2014})},\ \Eprint
  {http://arxiv.org/abs/1402.6672} {arXiv:1402.6672 [astro-ph.HE]} \BibitemShut
  {NoStop}%
\bibitem [{\citenamefont {Portegies~Zwart}\ \emph {et~al.}(2004)\citenamefont
  {Portegies~Zwart}, \citenamefont {Baumgardt}, \citenamefont {Hut},
  \citenamefont {Makino},\ and\ \citenamefont
  {McMillan}}]{PortegiesZwart:2004ggg}%
  \BibitemOpen
  \bibfield  {author} {\bibinfo {author} {\bibfnamefont {S.~F.}\ \bibnamefont
  {Portegies~Zwart}}, \bibinfo {author} {\bibfnamefont {H.}~\bibnamefont
  {Baumgardt}}, \bibinfo {author} {\bibfnamefont {P.}~\bibnamefont {Hut}},
  \bibinfo {author} {\bibfnamefont {J.}~\bibnamefont {Makino}}, \ and\ \bibinfo
  {author} {\bibfnamefont {S.~L.~W.}\ \bibnamefont {McMillan}},\ }\href
  {\doibase 10.1038/nature02448} {\bibfield  {journal} {\bibinfo  {journal}
  {Nature}\ }\textbf {\bibinfo {volume} {428}},\ \bibinfo {pages} {724}
  (\bibinfo {year} {2004})},\ \Eprint {http://arxiv.org/abs/astro-ph/0402622}
  {arXiv:astro-ph/0402622 [astro-ph]} \BibitemShut {NoStop}%
\bibitem [{\citenamefont {Miller}\ and\ \citenamefont
  {Hamilton}(2002)}]{Miller:2001ez}%
  \BibitemOpen
  \bibfield  {author} {\bibinfo {author} {\bibfnamefont {M.~C.}\ \bibnamefont
  {Miller}}\ and\ \bibinfo {author} {\bibfnamefont {D.~P.}\ \bibnamefont
  {Hamilton}},\ }\href {\doibase 10.1046/j.1365-8711.2002.05112.x} {\bibfield
  {journal} {\bibinfo  {journal} {Mon. Not. Roy. Astron. Soc.}\ }\textbf
  {\bibinfo {volume} {330}},\ \bibinfo {pages} {232} (\bibinfo {year}
  {2002})},\ \Eprint {http://arxiv.org/abs/astro-ph/0106188}
  {arXiv:astro-ph/0106188 [astro-ph]} \BibitemShut {NoStop}%
\bibitem [{\citenamefont {Portegies~Zwart}\ and\ \citenamefont
  {McMillan}(2002)}]{PortegiesZwart:2002iks}%
  \BibitemOpen
  \bibfield  {author} {\bibinfo {author} {\bibfnamefont {S.~F.}\ \bibnamefont
  {Portegies~Zwart}}\ and\ \bibinfo {author} {\bibfnamefont {S.~L.~W.}\
  \bibnamefont {McMillan}},\ }\href {\doibase 10.1086/341798} {\bibfield
  {journal} {\bibinfo  {journal} {Astrophys. J.}\ }\textbf {\bibinfo {volume}
  {576}},\ \bibinfo {pages} {899} (\bibinfo {year} {2002})},\ \Eprint
  {http://arxiv.org/abs/astro-ph/0201055} {arXiv:astro-ph/0201055 [astro-ph]}
  \BibitemShut {NoStop}%
\bibitem [{\citenamefont {Mapelli}(2016)}]{Mapelli:2016vca}%
  \BibitemOpen
  \bibfield  {author} {\bibinfo {author} {\bibfnamefont {M.}~\bibnamefont
  {Mapelli}},\ }\href {\doibase 10.1093/mnras/stw869} {\bibfield  {journal}
  {\bibinfo  {journal} {Mon. Not. Roy. Astron. Soc.}\ }\textbf {\bibinfo
  {volume} {459}},\ \bibinfo {pages} {3432} (\bibinfo {year} {2016})},\ \Eprint
  {http://arxiv.org/abs/1604.03559} {arXiv:1604.03559 [astro-ph.GA]}
  \BibitemShut {NoStop}%
\bibitem [{\citenamefont {Leigh}\ \emph {et~al.}(2014)\citenamefont {Leigh},
  \citenamefont {L{\"u}tzgendorf}, \citenamefont {Geller}, \citenamefont
  {Maccarone}, \citenamefont {Heinke},\ and\ \citenamefont
  {Sesana}}]{Leigh:2014oda}%
  \BibitemOpen
  \bibfield  {author} {\bibinfo {author} {\bibfnamefont {N.~W.~C.}\
  \bibnamefont {Leigh}}, \bibinfo {author} {\bibfnamefont {N.}~\bibnamefont
  {L{\"u}tzgendorf}}, \bibinfo {author} {\bibfnamefont {A.~M.}\ \bibnamefont
  {Geller}}, \bibinfo {author} {\bibfnamefont {T.~J.}\ \bibnamefont
  {Maccarone}}, \bibinfo {author} {\bibfnamefont {C.}~\bibnamefont {Heinke}}, \
  and\ \bibinfo {author} {\bibfnamefont {A.}~\bibnamefont {Sesana}},\ }\href
  {\doibase 10.1093/mnras/stu1437} {\bibfield  {journal} {\bibinfo  {journal}
  {Mon. Not. Roy. Astron. Soc.}\ }\textbf {\bibinfo {volume} {444}},\ \bibinfo
  {pages} {29} (\bibinfo {year} {2014})},\ \Eprint
  {http://arxiv.org/abs/1407.4459} {arXiv:1407.4459 [astro-ph.SR]} \BibitemShut
  {NoStop}%
\bibitem [{\citenamefont {Marchant}\ \emph {et~al.}(2016)\citenamefont
  {Marchant}, \citenamefont {Langer}, \citenamefont {Podsiadlowski},
  \citenamefont {Tauris},\ and\ \citenamefont {Moriya}}]{Marchant:2016wow}%
  \BibitemOpen
  \bibfield  {author} {\bibinfo {author} {\bibfnamefont {P.}~\bibnamefont
  {Marchant}}, \bibinfo {author} {\bibfnamefont {N.}~\bibnamefont {Langer}},
  \bibinfo {author} {\bibfnamefont {P.}~\bibnamefont {Podsiadlowski}}, \bibinfo
  {author} {\bibfnamefont {T.~M.}\ \bibnamefont {Tauris}}, \ and\ \bibinfo
  {author} {\bibfnamefont {T.~J.}\ \bibnamefont {Moriya}},\ }\href {\doibase
  10.1051/0004-6361/201628133} {\bibfield  {journal} {\bibinfo  {journal}
  {Astron. Astrophys.}\ }\textbf {\bibinfo {volume} {588}},\ \bibinfo {pages}
  {A50} (\bibinfo {year} {2016})},\ \Eprint {http://arxiv.org/abs/1601.03718}
  {arXiv:1601.03718 [astro-ph.SR]} \BibitemShut {NoStop}%
\bibitem [{\citenamefont {King}\ and\ \citenamefont
  {Dehnen}(2005)}]{King:2004ri}%
  \BibitemOpen
  \bibfield  {author} {\bibinfo {author} {\bibfnamefont {A.~R.}\ \bibnamefont
  {King}}\ and\ \bibinfo {author} {\bibfnamefont {W.}~\bibnamefont {Dehnen}},\
  }\href {\doibase 10.1111/j.1365-2966.2005.08634.x} {\bibfield  {journal}
  {\bibinfo  {journal} {Mon. Not. Roy. Astron. Soc.}\ }\textbf {\bibinfo
  {volume} {357}},\ \bibinfo {pages} {275} (\bibinfo {year} {2005})},\ \Eprint
  {http://arxiv.org/abs/astro-ph/0411489} {arXiv:astro-ph/0411489 [astro-ph]}
  \BibitemShut {NoStop}%
\bibitem [{\citenamefont {Abbott}\ \emph
  {et~al.}(2017{\natexlab{e}})\citenamefont {Abbott} \emph
  {et~al.}}]{Abbott:2017iws}%
  \BibitemOpen
  \bibfield  {author} {\bibinfo {author} {\bibfnamefont {B.~P.}\ \bibnamefont
  {Abbott}} \emph {et~al.} (\bibinfo {collaboration} {LIGO Scientific
  Collaboration, Virgo Collaboration}),\ }\href {\doibase
  10.1103/PhysRevD.96.022001} {\bibfield  {journal} {\bibinfo  {journal} {Phys.
  Rev.}\ }\textbf {\bibinfo {volume} {D96}},\ \bibinfo {pages} {022001}
  (\bibinfo {year} {2017}{\natexlab{e}})},\ \Eprint
  {http://arxiv.org/abs/1704.04628} {arXiv:1704.04628 [gr-qc]} \BibitemShut
  {NoStop}%
\bibitem [{\citenamefont {Abbott}\ \emph
  {et~al.}(2016{\natexlab{d}})\citenamefont {Abbott} \emph
  {et~al.}}]{TheLIGOScientific:2016src}%
  \BibitemOpen
  \bibfield  {author} {\bibinfo {author} {\bibfnamefont {B.~P.}\ \bibnamefont
  {Abbott}} \emph {et~al.} (\bibinfo {collaboration} {LIGO Scientific
  Collaboration, Virgo Collaboration}),\ }\href {\doibase
  10.1103/PhysRevLett.116.221101, 10.1103/PhysRevLett.121.129902} {\bibfield
  {journal} {\bibinfo  {journal} {Phys. Rev. Lett.}\ }\textbf {\bibinfo
  {volume} {116}},\ \bibinfo {pages} {221101} (\bibinfo {year}
  {2016}{\natexlab{d}})},\ \bibinfo {note} {[Erratum: Phys. Rev.
  Lett.121,no.12,129902(2018)]},\ \Eprint {http://arxiv.org/abs/1602.03841}
  {arXiv:1602.03841 [gr-qc]} \BibitemShut {NoStop}%
\bibitem [{\citenamefont {Yunes}\ \emph {et~al.}(2016)\citenamefont {Yunes},
  \citenamefont {Yagi},\ and\ \citenamefont {Pretorius}}]{Yunes:2016jcc}%
  \BibitemOpen
  \bibfield  {author} {\bibinfo {author} {\bibfnamefont {N.}~\bibnamefont
  {Yunes}}, \bibinfo {author} {\bibfnamefont {K.}~\bibnamefont {Yagi}}, \ and\
  \bibinfo {author} {\bibfnamefont {F.}~\bibnamefont {Pretorius}},\ }\href
  {\doibase 10.1103/PhysRevD.94.084002} {\bibfield  {journal} {\bibinfo
  {journal} {Phys. Rev.}\ }\textbf {\bibinfo {volume} {D94}},\ \bibinfo {pages}
  {084002} (\bibinfo {year} {2016})},\ \Eprint
  {http://arxiv.org/abs/1603.08955} {arXiv:1603.08955 [gr-qc]} \BibitemShut
  {NoStop}%
\bibitem [{\citenamefont {Yunes}\ and\ \citenamefont
  {Siemens}(2013)}]{Yunes:2013dva}%
  \BibitemOpen
  \bibfield  {author} {\bibinfo {author} {\bibfnamefont {N.}~\bibnamefont
  {Yunes}}\ and\ \bibinfo {author} {\bibfnamefont {X.}~\bibnamefont
  {Siemens}},\ }\href {\doibase 10.12942/lrr-2013-9} {\bibfield  {journal}
  {\bibinfo  {journal} {Living Rev. Rel.}\ }\textbf {\bibinfo {volume} {16}},\
  \bibinfo {pages} {9} (\bibinfo {year} {2013})},\ \Eprint
  {http://arxiv.org/abs/1304.3473} {arXiv:1304.3473 [gr-qc]} \BibitemShut
  {NoStop}%
\bibitem [{\citenamefont {Gair}\ \emph {et~al.}(2013)\citenamefont {Gair},
  \citenamefont {Vallisneri}, \citenamefont {Larson},\ and\ \citenamefont
  {Baker}}]{Gair:2012nm}%
  \BibitemOpen
  \bibfield  {author} {\bibinfo {author} {\bibfnamefont {J.~R.}\ \bibnamefont
  {Gair}}, \bibinfo {author} {\bibfnamefont {M.}~\bibnamefont {Vallisneri}},
  \bibinfo {author} {\bibfnamefont {S.~L.}\ \bibnamefont {Larson}}, \ and\
  \bibinfo {author} {\bibfnamefont {J.~G.}\ \bibnamefont {Baker}},\ }\href
  {\doibase 10.12942/lrr-2013-7} {\bibfield  {journal} {\bibinfo  {journal}
  {Living Rev. Rel.}\ }\textbf {\bibinfo {volume} {16}},\ \bibinfo {pages} {7}
  (\bibinfo {year} {2013})},\ \Eprint {http://arxiv.org/abs/1212.5575}
  {arXiv:1212.5575 [gr-qc]} \BibitemShut {NoStop}%
\bibitem [{\citenamefont {Kamaretsos}\ \emph {et~al.}(2012)\citenamefont
  {Kamaretsos}, \citenamefont {Hannam}, \citenamefont {Husa},\ and\
  \citenamefont {Sathyaprakash}}]{Kamaretsos:2011um}%
  \BibitemOpen
  \bibfield  {author} {\bibinfo {author} {\bibfnamefont {I.}~\bibnamefont
  {Kamaretsos}}, \bibinfo {author} {\bibfnamefont {M.}~\bibnamefont {Hannam}},
  \bibinfo {author} {\bibfnamefont {S.}~\bibnamefont {Husa}}, \ and\ \bibinfo
  {author} {\bibfnamefont {B.~S.}\ \bibnamefont {Sathyaprakash}},\ }\href
  {\doibase 10.1103/PhysRevD.85.024018} {\bibfield  {journal} {\bibinfo
  {journal} {Phys. Rev.}\ }\textbf {\bibinfo {volume} {D85}},\ \bibinfo {pages}
  {024018} (\bibinfo {year} {2012})},\ \Eprint {http://arxiv.org/abs/1107.0854}
  {arXiv:1107.0854 [gr-qc]} \BibitemShut {NoStop}%
\bibitem [{\citenamefont {Meidam}\ \emph {et~al.}(2014)\citenamefont {Meidam},
  \citenamefont {Agathos}, \citenamefont {Van Den~Broeck}, \citenamefont
  {Veitch},\ and\ \citenamefont {Sathyaprakash}}]{Meidam:2014jpa}%
  \BibitemOpen
  \bibfield  {author} {\bibinfo {author} {\bibfnamefont {J.}~\bibnamefont
  {Meidam}}, \bibinfo {author} {\bibfnamefont {M.}~\bibnamefont {Agathos}},
  \bibinfo {author} {\bibfnamefont {C.}~\bibnamefont {Van Den~Broeck}},
  \bibinfo {author} {\bibfnamefont {J.}~\bibnamefont {Veitch}}, \ and\ \bibinfo
  {author} {\bibfnamefont {B.~S.}\ \bibnamefont {Sathyaprakash}},\ }\href
  {\doibase 10.1103/PhysRevD.90.064009} {\bibfield  {journal} {\bibinfo
  {journal} {Phys. Rev.}\ }\textbf {\bibinfo {volume} {D90}},\ \bibinfo {pages}
  {064009} (\bibinfo {year} {2014})},\ \Eprint {http://arxiv.org/abs/1406.3201}
  {arXiv:1406.3201 [gr-qc]} \BibitemShut {NoStop}%
\bibitem [{\citenamefont {Thrane}\ \emph {et~al.}(2017)\citenamefont {Thrane},
  \citenamefont {Lasky},\ and\ \citenamefont {Levin}}]{Thrane:2017lqn}%
  \BibitemOpen
  \bibfield  {author} {\bibinfo {author} {\bibfnamefont {E.}~\bibnamefont
  {Thrane}}, \bibinfo {author} {\bibfnamefont {P.~D.}\ \bibnamefont {Lasky}}, \
  and\ \bibinfo {author} {\bibfnamefont {Y.}~\bibnamefont {Levin}},\ }\href
  {\doibase 10.1103/PhysRevD.96.102004} {\bibfield  {journal} {\bibinfo
  {journal} {Phys. Rev.}\ }\textbf {\bibinfo {volume} {D96}},\ \bibinfo {pages}
  {102004} (\bibinfo {year} {2017})},\ \Eprint
  {http://arxiv.org/abs/1706.05152} {arXiv:1706.05152 [gr-qc]} \BibitemShut
  {NoStop}%
\bibitem [{\citenamefont {Carullo}\ \emph {et~al.}(2018)\citenamefont {Carullo}
  \emph {et~al.}}]{Carullo:2018sfu}%
  \BibitemOpen
  \bibfield  {author} {\bibinfo {author} {\bibfnamefont {G.}~\bibnamefont
  {Carullo}} \emph {et~al.},\ }\href {\doibase 10.1103/PhysRevD.98.104020}
  {\bibfield  {journal} {\bibinfo  {journal} {Phys. Rev.}\ }\textbf {\bibinfo
  {volume} {D98}},\ \bibinfo {pages} {104020} (\bibinfo {year} {2018})},\
  \Eprint {http://arxiv.org/abs/1805.04760} {arXiv:1805.04760 [gr-qc]}
  \BibitemShut {NoStop}%
\bibitem [{\citenamefont {Gonzalez}\ \emph {et~al.}(2007)\citenamefont
  {Gonzalez}, \citenamefont {Sperhake}, \citenamefont {Bruegmann},
  \citenamefont {Hannam},\ and\ \citenamefont {Husa}}]{Gonzalez:2006md}%
  \BibitemOpen
  \bibfield  {author} {\bibinfo {author} {\bibfnamefont {J.~A.}\ \bibnamefont
  {Gonzalez}}, \bibinfo {author} {\bibfnamefont {U.}~\bibnamefont {Sperhake}},
  \bibinfo {author} {\bibfnamefont {B.}~\bibnamefont {Bruegmann}}, \bibinfo
  {author} {\bibfnamefont {M.}~\bibnamefont {Hannam}}, \ and\ \bibinfo {author}
  {\bibfnamefont {S.}~\bibnamefont {Husa}},\ }\href {\doibase
  10.1103/PhysRevLett.98.091101} {\bibfield  {journal} {\bibinfo  {journal}
  {Phys. Rev. Lett.}\ }\textbf {\bibinfo {volume} {98}},\ \bibinfo {pages}
  {091101} (\bibinfo {year} {2007})},\ \Eprint
  {http://arxiv.org/abs/gr-qc/0610154} {arXiv:gr-qc/0610154 [gr-qc]}
  \BibitemShut {NoStop}%
\bibitem [{\citenamefont {Campanelli}\ \emph {et~al.}(2007)\citenamefont
  {Campanelli}, \citenamefont {Lousto}, \citenamefont {Zlochower},\ and\
  \citenamefont {Merritt}}]{Campanelli:2007cga}%
  \BibitemOpen
  \bibfield  {author} {\bibinfo {author} {\bibfnamefont {M.}~\bibnamefont
  {Campanelli}}, \bibinfo {author} {\bibfnamefont {C.~O.}\ \bibnamefont
  {Lousto}}, \bibinfo {author} {\bibfnamefont {Y.}~\bibnamefont {Zlochower}}, \
  and\ \bibinfo {author} {\bibfnamefont {D.}~\bibnamefont {Merritt}},\ }\href
  {\doibase 10.1103/PhysRevLett.98.231102} {\bibfield  {journal} {\bibinfo
  {journal} {Phys. Rev. Lett.}\ }\textbf {\bibinfo {volume} {98}},\ \bibinfo
  {pages} {231102} (\bibinfo {year} {2007})},\ \Eprint
  {http://arxiv.org/abs/gr-qc/0702133} {arXiv:gr-qc/0702133 [GR-QC]}
  \BibitemShut {NoStop}%
\bibitem [{\citenamefont {Calder{\'o}n~Bustillo}\ \emph
  {et~al.}(2018)\citenamefont {Calder{\'o}n~Bustillo}, \citenamefont {Clark},
  \citenamefont {Laguna},\ and\ \citenamefont
  {Shoemaker}}]{CalderonBustillo:2018zuq}%
  \BibitemOpen
  \bibfield  {author} {\bibinfo {author} {\bibfnamefont {J.}~\bibnamefont
  {Calder{\'o}n~Bustillo}}, \bibinfo {author} {\bibfnamefont {J.~A.}\
  \bibnamefont {Clark}}, \bibinfo {author} {\bibfnamefont {P.}~\bibnamefont
  {Laguna}}, \ and\ \bibinfo {author} {\bibfnamefont {D.}~\bibnamefont
  {Shoemaker}},\ }\href {\doibase 10.1103/PhysRevLett.121.191102} {\bibfield
  {journal} {\bibinfo  {journal} {Phys. Rev. Lett.}\ }\textbf {\bibinfo
  {volume} {121}},\ \bibinfo {pages} {191102} (\bibinfo {year} {2018})},\
  \Eprint {http://arxiv.org/abs/1806.11160} {arXiv:1806.11160 [gr-qc]}
  \BibitemShut {NoStop}%
\bibitem [{\citenamefont {Abadie}\ \emph {et~al.}(2012)\citenamefont {Abadie}
  \emph {et~al.}}]{Virgo:2012aa}%
  \BibitemOpen
  \bibfield  {author} {\bibinfo {author} {\bibfnamefont {J.}~\bibnamefont
  {Abadie}} \emph {et~al.} (\bibinfo {collaboration} {LIGO Scientific
  Collaboration, Virgo Collaboration}),\ }\href {\doibase
  10.1103/PhysRevD.85.102004} {\bibfield  {journal} {\bibinfo  {journal} {Phys.
  Rev.}\ }\textbf {\bibinfo {volume} {D85}},\ \bibinfo {pages} {102004}
  (\bibinfo {year} {2012})},\ \Eprint {http://arxiv.org/abs/1201.5999}
  {arXiv:1201.5999 [gr-qc]} \BibitemShut {NoStop}%
\bibitem [{\citenamefont {Aasi}\ \emph {et~al.}(2014)\citenamefont {Aasi} \emph
  {et~al.}}]{Aasi:2014iwa}%
  \BibitemOpen
  \bibfield  {author} {\bibinfo {author} {\bibfnamefont {J.}~\bibnamefont
  {Aasi}} \emph {et~al.} (\bibinfo {collaboration} {LIGO Scientific
  Collaboration and Virgo Collaboration}),\ }\href {\doibase
  10.1103/PhysRevD.89.122003} {\bibfield  {journal} {\bibinfo  {journal} {Phys.
  Rev.}\ }\textbf {\bibinfo {volume} {D89}},\ \bibinfo {pages} {122003}
  (\bibinfo {year} {2014})},\ \Eprint {http://arxiv.org/abs/1404.2199}
  {arXiv:1404.2199 [gr-qc]} \BibitemShut {NoStop}%
\bibitem [{\citenamefont {Klimenko}\ \emph {et~al.}(2005)\citenamefont
  {Klimenko}, \citenamefont {Mohanty}, \citenamefont {Rakhmanov},\ and\
  \citenamefont {Mitselmakher}}]{klimenko:2005xv}%
  \BibitemOpen
  \bibfield  {author} {\bibinfo {author} {\bibfnamefont {S.}~\bibnamefont
  {Klimenko}}, \bibinfo {author} {\bibfnamefont {S.}~\bibnamefont {Mohanty}},
  \bibinfo {author} {\bibfnamefont {M.}~\bibnamefont {Rakhmanov}}, \ and\
  \bibinfo {author} {\bibfnamefont {G.}~\bibnamefont {Mitselmakher}},\ }\href
  {\doibase 10.1103/PhysRevD.72.122002} {\bibfield  {journal} {\bibinfo
  {journal} {Phys. Rev.}\ }\textbf {\bibinfo {volume} {D72}},\ \bibinfo {pages}
  {122002} (\bibinfo {year} {2005})},\ \Eprint
  {http://arxiv.org/abs/gr-qc/0508068} {arXiv:gr-qc/0508068 [gr-qc]}
  \BibitemShut {NoStop}%
\bibitem [{\citenamefont {Aasi}\ \emph {et~al.}(2015)\citenamefont {Aasi} \emph
  {et~al.}}]{TheLIGOScientific:2014jea}%
  \BibitemOpen
  \bibfield  {author} {\bibinfo {author} {\bibfnamefont {J.}~\bibnamefont
  {Aasi}} \emph {et~al.} (\bibinfo {collaboration} {LIGO Scientific
  Collaboration}),\ }\href {\doibase 10.1088/0264-9381/32/7/074001} {\bibfield
  {journal} {\bibinfo  {journal} {Class. Quant. Grav.}\ }\textbf {\bibinfo
  {volume} {32}},\ \bibinfo {pages} {074001} (\bibinfo {year} {2015})},\
  \Eprint {http://arxiv.org/abs/1411.4547} {arXiv:1411.4547 [gr-qc]}
  \BibitemShut {NoStop}%
\bibitem [{\citenamefont {Acernese}\ \emph {et~al.}(2015)\citenamefont
  {Acernese} \emph {et~al.}}]{TheVirgo:2014hva}%
  \BibitemOpen
  \bibfield  {author} {\bibinfo {author} {\bibfnamefont {F.}~\bibnamefont
  {Acernese}} \emph {et~al.} (\bibinfo {collaboration} {VIRGO}),\ }\href
  {\doibase 10.1088/0264-9381/32/2/024001} {\bibfield  {journal} {\bibinfo
  {journal} {Class. Quant. Grav.}\ }\textbf {\bibinfo {volume} {32}},\ \bibinfo
  {pages} {024001} (\bibinfo {year} {2015})},\ \Eprint
  {http://arxiv.org/abs/1408.3978} {arXiv:1408.3978 [gr-qc]} \BibitemShut
  {NoStop}%
\bibitem [{\citenamefont {Cannon}\ \emph {et~al.}(2012)\citenamefont {Cannon}
  \emph {et~al.}}]{Cannon:2011vi}%
  \BibitemOpen
  \bibfield  {author} {\bibinfo {author} {\bibfnamefont {K.}~\bibnamefont
  {Cannon}} \emph {et~al.},\ }\href {\doibase 10.1088/0004-637X/748/2/136}
  {\bibfield  {journal} {\bibinfo  {journal} {Astrophys. J.}\ }\textbf
  {\bibinfo {volume} {748}},\ \bibinfo {pages} {136} (\bibinfo {year}
  {2012})},\ \Eprint {http://arxiv.org/abs/1107.2665} {arXiv:1107.2665
  [astro-ph.IM]} \BibitemShut {NoStop}%
\bibitem [{\citenamefont {Privitera}\ \emph {et~al.}(2014)\citenamefont
  {Privitera}, \citenamefont {Mohapatra}, \citenamefont {Ajith}, \citenamefont
  {Cannon}, \citenamefont {Fotopoulos}, \citenamefont {Frei}, \citenamefont
  {Hanna}, \citenamefont {Weinstein},\ and\ \citenamefont
  {Whelan}}]{Privitera:2013xza}%
  \BibitemOpen
  \bibfield  {author} {\bibinfo {author} {\bibfnamefont {S.}~\bibnamefont
  {Privitera}}, \bibinfo {author} {\bibfnamefont {S.~R.~P.}\ \bibnamefont
  {Mohapatra}}, \bibinfo {author} {\bibfnamefont {P.}~\bibnamefont {Ajith}},
  \bibinfo {author} {\bibfnamefont {K.}~\bibnamefont {Cannon}}, \bibinfo
  {author} {\bibfnamefont {N.}~\bibnamefont {Fotopoulos}}, \bibinfo {author}
  {\bibfnamefont {M.~A.}\ \bibnamefont {Frei}}, \bibinfo {author}
  {\bibfnamefont {C.}~\bibnamefont {Hanna}}, \bibinfo {author} {\bibfnamefont
  {A.~J.}\ \bibnamefont {Weinstein}}, \ and\ \bibinfo {author} {\bibfnamefont
  {J.~T.}\ \bibnamefont {Whelan}},\ }\href {\doibase
  10.1103/PhysRevD.89.024003} {\bibfield  {journal} {\bibinfo  {journal} {Phys.
  Rev.}\ }\textbf {\bibinfo {volume} {D89}},\ \bibinfo {pages} {024003}
  (\bibinfo {year} {2014})},\ \Eprint {http://arxiv.org/abs/1310.5633}
  {arXiv:1310.5633 [gr-qc]} \BibitemShut {NoStop}%
\bibitem [{\citenamefont {Messick}\ \emph {et~al.}(2017)\citenamefont {Messick}
  \emph {et~al.}}]{Messick:2016aqy}%
  \BibitemOpen
  \bibfield  {author} {\bibinfo {author} {\bibfnamefont {C.}~\bibnamefont
  {Messick}} \emph {et~al.},\ }\href {\doibase 10.1103/PhysRevD.95.042001}
  {\bibfield  {journal} {\bibinfo  {journal} {Phys. Rev.}\ }\textbf {\bibinfo
  {volume} {D95}},\ \bibinfo {pages} {042001} (\bibinfo {year} {2017})},\
  \Eprint {http://arxiv.org/abs/1604.04324} {arXiv:1604.04324 [astro-ph.IM]}
  \BibitemShut {NoStop}%
\bibitem [{\citenamefont {Taracchini}\ \emph {et~al.}(2012)\citenamefont
  {Taracchini}, \citenamefont {Pan}, \citenamefont {Buonanno}, \citenamefont
  {Barausse}, \citenamefont {Boyle}, \citenamefont {Chu}, \citenamefont
  {Lovelace}, \citenamefont {Pfeiffer},\ and\ \citenamefont
  {Scheel}}]{Taracchini:2012ig}%
  \BibitemOpen
  \bibfield  {author} {\bibinfo {author} {\bibfnamefont {A.}~\bibnamefont
  {Taracchini}}, \bibinfo {author} {\bibfnamefont {Y.}~\bibnamefont {Pan}},
  \bibinfo {author} {\bibfnamefont {A.}~\bibnamefont {Buonanno}}, \bibinfo
  {author} {\bibfnamefont {E.}~\bibnamefont {Barausse}}, \bibinfo {author}
  {\bibfnamefont {M.}~\bibnamefont {Boyle}}, \bibinfo {author} {\bibfnamefont
  {T.}~\bibnamefont {Chu}}, \bibinfo {author} {\bibfnamefont {G.}~\bibnamefont
  {Lovelace}}, \bibinfo {author} {\bibfnamefont {H.~P.}\ \bibnamefont
  {Pfeiffer}}, \ and\ \bibinfo {author} {\bibfnamefont {M.~A.}\ \bibnamefont
  {Scheel}},\ }\href {\doibase 10.1103/PhysRevD.86.024011} {\bibfield
  {journal} {\bibinfo  {journal} {Phys. Rev.}\ }\textbf {\bibinfo {volume}
  {D86}},\ \bibinfo {pages} {024011} (\bibinfo {year} {2012})},\ \Eprint
  {http://arxiv.org/abs/1202.0790} {arXiv:1202.0790 [gr-qc]} \BibitemShut
  {NoStop}%
\bibitem [{\citenamefont {Pekowsky}\ \emph {et~al.}(2013)\citenamefont
  {Pekowsky}, \citenamefont {Healy}, \citenamefont {Shoemaker},\ and\
  \citenamefont {Laguna}}]{Pekowsky:2012sr}%
  \BibitemOpen
  \bibfield  {author} {\bibinfo {author} {\bibfnamefont {L.}~\bibnamefont
  {Pekowsky}}, \bibinfo {author} {\bibfnamefont {J.}~\bibnamefont {Healy}},
  \bibinfo {author} {\bibfnamefont {D.}~\bibnamefont {Shoemaker}}, \ and\
  \bibinfo {author} {\bibfnamefont {P.}~\bibnamefont {Laguna}},\ }\href
  {\doibase 10.1103/PhysRevD.87.084008} {\bibfield  {journal} {\bibinfo
  {journal} {Phys.Rev.}\ }\textbf {\bibinfo {volume} {D87}},\ \bibinfo {pages}
  {084008} (\bibinfo {year} {2013})},\ \Eprint {http://arxiv.org/abs/1210.1891}
  {arXiv:1210.1891 [gr-qc]} \BibitemShut {NoStop}%
\bibitem [{\citenamefont {Varma}\ \emph {et~al.}(2014)\citenamefont {Varma},
  \citenamefont {Ajith}, \citenamefont {Husa}, \citenamefont {Bustillo},
  \citenamefont {Hannam},\ and\ \citenamefont {P\"urrer}}]{Varma:2014jxa}%
  \BibitemOpen
  \bibfield  {author} {\bibinfo {author} {\bibfnamefont {V.}~\bibnamefont
  {Varma}}, \bibinfo {author} {\bibfnamefont {P.}~\bibnamefont {Ajith}},
  \bibinfo {author} {\bibfnamefont {S.}~\bibnamefont {Husa}}, \bibinfo {author}
  {\bibfnamefont {J.~C.}\ \bibnamefont {Bustillo}}, \bibinfo {author}
  {\bibfnamefont {M.}~\bibnamefont {Hannam}}, \ and\ \bibinfo {author}
  {\bibfnamefont {M.}~\bibnamefont {P\"urrer}},\ }\href {\doibase
  10.1103/PhysRevD.90.124004} {\bibfield  {journal} {\bibinfo  {journal} {Phys.
  Rev.}\ }\textbf {\bibinfo {volume} {D90}},\ \bibinfo {pages} {124004}
  (\bibinfo {year} {2014})},\ \Eprint {http://arxiv.org/abs/1409.2349}
  {arXiv:1409.2349 [gr-qc]} \BibitemShut {NoStop}%
\bibitem [{\citenamefont {Calder{\'o}n~Bustillo}\ \emph
  {et~al.}(2016)\citenamefont {Calder{\'o}n~Bustillo}, \citenamefont {Husa},
  \citenamefont {Sintes},\ and\ \citenamefont {P{\"u}rrer}}]{Bustillo:2015qty}%
  \BibitemOpen
  \bibfield  {author} {\bibinfo {author} {\bibfnamefont {J.}~\bibnamefont
  {Calder{\'o}n~Bustillo}}, \bibinfo {author} {\bibfnamefont {S.}~\bibnamefont
  {Husa}}, \bibinfo {author} {\bibfnamefont {A.~M.}\ \bibnamefont {Sintes}}, \
  and\ \bibinfo {author} {\bibfnamefont {M.}~\bibnamefont {P{\"u}rrer}},\
  }\href {\doibase 10.1103/PhysRevD.93.084019} {\bibfield  {journal} {\bibinfo
  {journal} {Phys. Rev.}\ }\textbf {\bibinfo {volume} {D93}},\ \bibinfo {pages}
  {084019} (\bibinfo {year} {2016})},\ \Eprint
  {http://arxiv.org/abs/1511.02060} {arXiv:1511.02060 [gr-qc]} \BibitemShut
  {NoStop}%
\bibitem [{\citenamefont {Varma}\ and\ \citenamefont
  {Ajith}(2016)}]{Varma:2016dnf}%
  \BibitemOpen
  \bibfield  {author} {\bibinfo {author} {\bibfnamefont {V.}~\bibnamefont
  {Varma}}\ and\ \bibinfo {author} {\bibfnamefont {P.}~\bibnamefont {Ajith}},\
  }\href@noop {} {\  (\bibinfo {year} {2016})},\ \Eprint
  {http://arxiv.org/abs/1612.05608} {arXiv:1612.05608 [gr-qc]} \BibitemShut
  {NoStop}%
\bibitem [{\citenamefont {Calder{\'o}n~Bustillo}\ \emph
  {et~al.}(2017)\citenamefont {Calder{\'o}n~Bustillo}, \citenamefont {Laguna},\
  and\ \citenamefont {Shoemaker}}]{Bustillo:2016gid}%
  \BibitemOpen
  \bibfield  {author} {\bibinfo {author} {\bibfnamefont {J.}~\bibnamefont
  {Calder{\'o}n~Bustillo}}, \bibinfo {author} {\bibfnamefont {P.}~\bibnamefont
  {Laguna}}, \ and\ \bibinfo {author} {\bibfnamefont {D.}~\bibnamefont
  {Shoemaker}},\ }\href {\doibase 10.1103/PhysRevD.95.104038} {\bibfield
  {journal} {\bibinfo  {journal} {Phys. Rev.}\ }\textbf {\bibinfo {volume}
  {D95}},\ \bibinfo {pages} {104038} (\bibinfo {year} {2017})},\ \Eprint
  {http://arxiv.org/abs/1612.02340} {arXiv:1612.02340 [gr-qc]} \BibitemShut
  {NoStop}%
\bibitem [{\citenamefont {Calder\'{o}n~Bustillo}\ \emph
  {et~al.}(2017)\citenamefont {Calder\'{o}n~Bustillo}, \citenamefont {Salemi},
  \citenamefont {Dal~Canton},\ and\ \citenamefont
  {Jani}}]{CalderonBustillo:2017skv}%
  \BibitemOpen
  \bibfield  {author} {\bibinfo {author} {\bibfnamefont {J.}~\bibnamefont
  {Calder\'{o}n~Bustillo}}, \bibinfo {author} {\bibfnamefont {F.}~\bibnamefont
  {Salemi}}, \bibinfo {author} {\bibfnamefont {T.}~\bibnamefont {Dal~Canton}},
  \ and\ \bibinfo {author} {\bibfnamefont {K.}~\bibnamefont {Jani}},\
  }\href@noop {} {\  (\bibinfo {year} {2017})},\ \Eprint
  {http://arxiv.org/abs/1711.02009} {arXiv:1711.02009 [gr-qc]} \BibitemShut
  {NoStop}%
\bibitem [{\citenamefont {Dal~Canton}\ \emph
  {et~al.}(2014{\natexlab{a}})\citenamefont {Dal~Canton} \emph
  {et~al.}}]{Canton:2014ena}%
  \BibitemOpen
  \bibfield  {author} {\bibinfo {author} {\bibfnamefont {T.}~\bibnamefont
  {Dal~Canton}} \emph {et~al.},\ }\href {\doibase 10.1103/PhysRevD.90.082004}
  {\bibfield  {journal} {\bibinfo  {journal} {Phys. Rev.}\ }\textbf {\bibinfo
  {volume} {D90}},\ \bibinfo {pages} {082004} (\bibinfo {year}
  {2014}{\natexlab{a}})},\ \Eprint {http://arxiv.org/abs/1405.6731}
  {arXiv:1405.6731 [gr-qc]} \BibitemShut {NoStop}%
\bibitem [{\citenamefont {Usman}\ \emph {et~al.}(2016)\citenamefont {Usman}
  \emph {et~al.}}]{Usman:2015kfa}%
  \BibitemOpen
  \bibfield  {author} {\bibinfo {author} {\bibfnamefont {S.~A.}\ \bibnamefont
  {Usman}} \emph {et~al.},\ }\href {\doibase 10.1088/0264-9381/33/21/215004}
  {\bibfield  {journal} {\bibinfo  {journal} {Class. Quant. Grav.}\ }\textbf
  {\bibinfo {volume} {33}},\ \bibinfo {pages} {215004} (\bibinfo {year}
  {2016})},\ \Eprint {http://arxiv.org/abs/1508.02357} {arXiv:1508.02357
  [gr-qc]} \BibitemShut {NoStop}%
\bibitem [{\citenamefont {Cahillane}\ \emph {et~al.}(2017)\citenamefont
  {Cahillane} \emph {et~al.}}]{Cahillane:2017vkb}%
  \BibitemOpen
  \bibfield  {author} {\bibinfo {author} {\bibfnamefont {C.}~\bibnamefont
  {Cahillane}} \emph {et~al.} (\bibinfo {collaboration} {LIGO Scientific}),\
  }\href {\doibase 10.1103/PhysRevD.96.102001} {\bibfield  {journal} {\bibinfo
  {journal} {Phys. Rev.}\ }\textbf {\bibinfo {volume} {D96}},\ \bibinfo {pages}
  {102001} (\bibinfo {year} {2017})},\ \Eprint
  {http://arxiv.org/abs/1708.03023} {arXiv:1708.03023 [astro-ph.IM]}
  \BibitemShut {NoStop}%
\bibitem [{\citenamefont {Viets}\ \emph {et~al.}(2018)\citenamefont {Viets}
  \emph {et~al.}}]{Viets:2017yvy}%
  \BibitemOpen
  \bibfield  {author} {\bibinfo {author} {\bibfnamefont {A.}~\bibnamefont
  {Viets}} \emph {et~al.},\ }\href {\doibase 10.1088/1361-6382/aab658}
  {\bibfield  {journal} {\bibinfo  {journal} {Class. Quant. Grav.}\ }\textbf
  {\bibinfo {volume} {35}},\ \bibinfo {pages} {095015} (\bibinfo {year}
  {2018})},\ \Eprint {http://arxiv.org/abs/1710.09973} {arXiv:1710.09973
  [astro-ph.IM]} \BibitemShut {NoStop}%
\bibitem [{\citenamefont {Acernese}\ \emph {et~al.}(2018)\citenamefont
  {Acernese} \emph {et~al.}}]{Acernese:2018bfl}%
  \BibitemOpen
  \bibfield  {author} {\bibinfo {author} {\bibfnamefont {F.}~\bibnamefont
  {Acernese}} \emph {et~al.} (\bibinfo {collaboration} {Virgo}),\ }\href
  {\doibase 10.1088/1361-6382/aadf1a} {\bibfield  {journal} {\bibinfo
  {journal} {Class. Quant. Grav.}\ }\textbf {\bibinfo {volume} {35}},\ \bibinfo
  {pages} {205004} (\bibinfo {year} {2018})},\ \Eprint
  {http://arxiv.org/abs/1807.03275} {arXiv:1807.03275 [gr-qc]} \BibitemShut
  {NoStop}%
\bibitem [{\citenamefont {Driggers}\ \emph {et~al.}(2018)\citenamefont
  {Driggers} \emph {et~al.}}]{Driggers:2018gii}%
  \BibitemOpen
  \bibfield  {author} {\bibinfo {author} {\bibfnamefont {J.~C.}\ \bibnamefont
  {Driggers}} \emph {et~al.} (\bibinfo {collaboration} {LIGO Scientific
  Collaboration}),\ }\href@noop {} {\  (\bibinfo {year} {2018})},\ \Eprint
  {http://arxiv.org/abs/1806.00532} {1806.00532 [astro-ph.IM]} \BibitemShut
  {NoStop}%
\bibitem [{\citenamefont {Sachdev}\ \emph {et~al.}(2019)\citenamefont
  {Sachdev}, \citenamefont {Caudill}, \citenamefont {Fong}, \citenamefont {Lo},
  \citenamefont {Messick}, \citenamefont {Mukherjee}, \citenamefont {Magee},
  \citenamefont {Tsukada}, \citenamefont {Blackburn}, \citenamefont {Brady}
  \emph {et~al.}}]{sachdev2019gstlal}%
  \BibitemOpen
  \bibfield  {author} {\bibinfo {author} {\bibfnamefont {S.}~\bibnamefont
  {Sachdev}}, \bibinfo {author} {\bibfnamefont {S.}~\bibnamefont {Caudill}},
  \bibinfo {author} {\bibfnamefont {H.}~\bibnamefont {Fong}}, \bibinfo {author}
  {\bibfnamefont {R.~K.}\ \bibnamefont {Lo}}, \bibinfo {author} {\bibfnamefont
  {C.}~\bibnamefont {Messick}}, \bibinfo {author} {\bibfnamefont
  {D.}~\bibnamefont {Mukherjee}}, \bibinfo {author} {\bibfnamefont
  {R.}~\bibnamefont {Magee}}, \bibinfo {author} {\bibfnamefont
  {L.}~\bibnamefont {Tsukada}}, \bibinfo {author} {\bibfnamefont
  {K.}~\bibnamefont {Blackburn}}, \bibinfo {author} {\bibfnamefont
  {P.}~\bibnamefont {Brady}},  \emph {et~al.},\ }\href@noop {} {\bibfield
  {journal} {\bibinfo  {journal} {arXiv preprint arXiv:1901.08580}\ } (\bibinfo
  {year} {2019})}\BibitemShut {NoStop}%
\bibitem [{\citenamefont {Klimenko}\ \emph {et~al.}(2016)\citenamefont
  {Klimenko} \emph {et~al.}}]{Klimenko:2015ypf}%
  \BibitemOpen
  \bibfield  {author} {\bibinfo {author} {\bibfnamefont {S.}~\bibnamefont
  {Klimenko}} \emph {et~al.},\ }\href {\doibase 10.1103/PhysRevD.93.042004}
  {\bibfield  {journal} {\bibinfo  {journal} {Phys. Rev.}\ }\textbf {\bibinfo
  {volume} {D93}},\ \bibinfo {pages} {042004} (\bibinfo {year} {2016})},\
  \Eprint {http://arxiv.org/abs/1511.05999} {arXiv:1511.05999 [gr-qc]}
  \BibitemShut {NoStop}%
\bibitem [{\citenamefont {Wainstein}\ and\ \citenamefont
  {Zubakov}(1962)}]{MatchedFilter}%
  \BibitemOpen
  \bibfield  {author} {\bibinfo {author} {\bibfnamefont {L.}~\bibnamefont
  {Wainstein}}\ and\ \bibinfo {author} {\bibfnamefont {V.}~\bibnamefont
  {Zubakov}},\ }\href@noop {} {\bibfield  {journal} {\bibinfo  {journal}
  {Prentice-Hall, Englewood Cliffs}\ } (\bibinfo {year} {1962})}\BibitemShut
  {NoStop}%
\bibitem [{\citenamefont {Capano}\ \emph {et~al.}(2014)\citenamefont {Capano},
  \citenamefont {Pan},\ and\ \citenamefont {Buonanno}}]{Capano:2013raa}%
  \BibitemOpen
  \bibfield  {author} {\bibinfo {author} {\bibfnamefont {C.}~\bibnamefont
  {Capano}}, \bibinfo {author} {\bibfnamefont {Y.}~\bibnamefont {Pan}}, \ and\
  \bibinfo {author} {\bibfnamefont {A.}~\bibnamefont {Buonanno}},\ }\href
  {\doibase 10.1103/PhysRevD.89.102003} {\bibfield  {journal} {\bibinfo
  {journal} {Phys. Rev.}\ }\textbf {\bibinfo {volume} {D89}},\ \bibinfo {pages}
  {102003} (\bibinfo {year} {2014})},\ \Eprint {http://arxiv.org/abs/1311.1286}
  {arXiv:1311.1286 [gr-qc]} \BibitemShut {NoStop}%
\bibitem [{\citenamefont {Cabero}\ \emph {et~al.}(2019)\citenamefont {Cabero}
  \emph {et~al.}}]{Cabero:2019orq}%
  \BibitemOpen
  \bibfield  {author} {\bibinfo {author} {\bibfnamefont {M.}~\bibnamefont
  {Cabero}} \emph {et~al.},\ }\href@noop {} {\  (\bibinfo {year} {2019})},\
  \Eprint {http://arxiv.org/abs/1901.05093} {arXiv:1901.05093
  [physics.ins-det]} \BibitemShut {NoStop}%
\bibitem [{\citenamefont {Dal~Canton}\ \emph
  {et~al.}(2014{\natexlab{b}})\citenamefont {Dal~Canton}, \citenamefont
  {Bhagwat}, \citenamefont {Dhurandhar},\ and\ \citenamefont
  {Lundgren}}]{Canton:2013joa}%
  \BibitemOpen
  \bibfield  {author} {\bibinfo {author} {\bibfnamefont {T.}~\bibnamefont
  {Dal~Canton}}, \bibinfo {author} {\bibfnamefont {S.}~\bibnamefont {Bhagwat}},
  \bibinfo {author} {\bibfnamefont {S.}~\bibnamefont {Dhurandhar}}, \ and\
  \bibinfo {author} {\bibfnamefont {A.}~\bibnamefont {Lundgren}},\ }\href
  {\doibase 10.1088/0264-9381/31/1/015016} {\bibfield  {journal} {\bibinfo
  {journal} {Class.Quant.Grav.}\ }\textbf {\bibinfo {volume} {31}},\ \bibinfo
  {pages} {015016} (\bibinfo {year} {2014}{\natexlab{b}})},\ \Eprint
  {http://arxiv.org/abs/1304.0008} {arXiv:1304.0008 [gr-qc]} \BibitemShut
  {NoStop}%
\bibitem [{\citenamefont {Nitz}(2018)}]{Nitz:2017lco}%
  \BibitemOpen
  \bibfield  {author} {\bibinfo {author} {\bibfnamefont {A.~H.}\ \bibnamefont
  {Nitz}},\ }\href {\doibase 10.1088/1361-6382/aaa13d} {\bibfield  {journal}
  {\bibinfo  {journal} {Class. Quant. Grav.}\ }\textbf {\bibinfo {volume}
  {35}},\ \bibinfo {pages} {035016} (\bibinfo {year} {2018})},\ \Eprint
  {http://arxiv.org/abs/1709.08974} {arXiv:1709.08974 [gr-qc]} \BibitemShut
  {NoStop}%
\bibitem [{\citenamefont {Allen}(2005)}]{Allen:2004gu}%
  \BibitemOpen
  \bibfield  {author} {\bibinfo {author} {\bibfnamefont {B.}~\bibnamefont
  {Allen}},\ }\href {\doibase 10.1103/PhysRevD.71.062001} {\bibfield  {journal}
  {\bibinfo  {journal} {Phys. Rev.}\ }\textbf {\bibinfo {volume} {D71}},\
  \bibinfo {pages} {062001} (\bibinfo {year} {2005})},\ \Eprint
  {http://arxiv.org/abs/gr-qc/0405045} {arXiv:gr-qc/0405045 [gr-qc]}
  \BibitemShut {NoStop}%
\bibitem [{\citenamefont {{Mukherjee}}\ \emph {et~al.}(2018)\citenamefont
  {{Mukherjee}}, \citenamefont {{Caudill}}, \citenamefont {{Magee}},
  \citenamefont {{Messick}}, \citenamefont {{Privitera}}, \citenamefont
  {{Sachdev}}, \citenamefont {{Blackburn}}, \citenamefont {{Brady}},
  \citenamefont {{Brockill}}, \citenamefont {{Cannon}}, \citenamefont
  {{Chamberlin}}, \citenamefont {{Chatterjee}}, \citenamefont {{Creighton}},
  \citenamefont {{Fong}}, \citenamefont {{Godwin}}, \citenamefont {{Hanna}},
  \citenamefont {{Kapadia}}, \citenamefont {{Lang}}, \citenamefont {{Li}},
  \citenamefont {{Lo}}, \citenamefont {{Meacher}}, \citenamefont {{Pace}},
  \citenamefont {{Sadeghian}}, \citenamefont {{Tsukada}}, \citenamefont
  {{Wade}}, \citenamefont {{Wade}}, \citenamefont {{Weinstein}},\ and\
  \citenamefont {{Xiao}}}]{gstlal_bank}%
  \BibitemOpen
  \bibfield  {author} {\bibinfo {author} {\bibfnamefont {D.}~\bibnamefont
  {{Mukherjee}}}, \bibinfo {author} {\bibfnamefont {S.}~\bibnamefont
  {{Caudill}}}, \bibinfo {author} {\bibfnamefont {R.}~\bibnamefont {{Magee}}},
  \bibinfo {author} {\bibfnamefont {C.}~\bibnamefont {{Messick}}}, \bibinfo
  {author} {\bibfnamefont {S.}~\bibnamefont {{Privitera}}}, \bibinfo {author}
  {\bibfnamefont {S.}~\bibnamefont {{Sachdev}}}, \bibinfo {author}
  {\bibfnamefont {K.}~\bibnamefont {{Blackburn}}}, \bibinfo {author}
  {\bibfnamefont {P.}~\bibnamefont {{Brady}}}, \bibinfo {author} {\bibfnamefont
  {P.}~\bibnamefont {{Brockill}}}, \bibinfo {author} {\bibfnamefont
  {K.}~\bibnamefont {{Cannon}}}, \bibinfo {author} {\bibfnamefont {S.~J.}\
  \bibnamefont {{Chamberlin}}}, \bibinfo {author} {\bibfnamefont
  {D.}~\bibnamefont {{Chatterjee}}}, \bibinfo {author} {\bibfnamefont
  {J.~D.~E.}\ \bibnamefont {{Creighton}}}, \bibinfo {author} {\bibfnamefont
  {H.}~\bibnamefont {{Fong}}}, \bibinfo {author} {\bibfnamefont
  {P.}~\bibnamefont {{Godwin}}}, \bibinfo {author} {\bibfnamefont
  {C.}~\bibnamefont {{Hanna}}}, \bibinfo {author} {\bibfnamefont
  {S.}~\bibnamefont {{Kapadia}}}, \bibinfo {author} {\bibfnamefont {R.~N.}\
  \bibnamefont {{Lang}}}, \bibinfo {author} {\bibfnamefont {T.~G.~F.}\
  \bibnamefont {{Li}}}, \bibinfo {author} {\bibfnamefont {R.~K.~L.}\
  \bibnamefont {{Lo}}}, \bibinfo {author} {\bibfnamefont {D.}~\bibnamefont
  {{Meacher}}}, \bibinfo {author} {\bibfnamefont {A.}~\bibnamefont {{Pace}}},
  \bibinfo {author} {\bibfnamefont {L.}~\bibnamefont {{Sadeghian}}}, \bibinfo
  {author} {\bibfnamefont {L.}~\bibnamefont {{Tsukada}}}, \bibinfo {author}
  {\bibfnamefont {L.}~\bibnamefont {{Wade}}}, \bibinfo {author} {\bibfnamefont
  {M.}~\bibnamefont {{Wade}}}, \bibinfo {author} {\bibfnamefont
  {A.}~\bibnamefont {{Weinstein}}}, \ and\ \bibinfo {author} {\bibfnamefont
  {L.}~\bibnamefont {{Xiao}}},\ }\href@noop {} {\bibfield  {journal} {\bibinfo
  {journal} {arXiv e-prints}\ ,\ \bibinfo {eid} {arXiv:1812.05121}} (\bibinfo
  {year} {2018})},\ \Eprint {http://arxiv.org/abs/1812.05121} {arXiv:1812.05121
  [astro-ph.IM]} \BibitemShut {NoStop}%
\bibitem [{\citenamefont {Boh{\'e}}\ \emph {et~al.}(2017)\citenamefont
  {Boh{\'e}} \emph {et~al.}}]{Bohe:2016gbl}%
  \BibitemOpen
  \bibfield  {author} {\bibinfo {author} {\bibfnamefont {A.}~\bibnamefont
  {Boh{\'e}}} \emph {et~al.},\ }\href {\doibase 10.1103/PhysRevD.95.044028}
  {\bibfield  {journal} {\bibinfo  {journal} {Phys. Rev.}\ }\textbf {\bibinfo
  {volume} {D95}},\ \bibinfo {pages} {044028} (\bibinfo {year} {2017})},\
  \Eprint {http://arxiv.org/abs/1611.03703} {arXiv:1611.03703 [gr-qc]}
  \BibitemShut {NoStop}%
\bibitem [{\citenamefont {Dal~Canton}\ and\ \citenamefont
  {Harry}(2017)}]{DalCanton:2017ala}%
  \BibitemOpen
  \bibfield  {author} {\bibinfo {author} {\bibfnamefont {T.}~\bibnamefont
  {Dal~Canton}}\ and\ \bibinfo {author} {\bibfnamefont {I.~W.}\ \bibnamefont
  {Harry}},\ }\href@noop {} {\  (\bibinfo {year} {2017})},\ \Eprint
  {http://arxiv.org/abs/1705.01845} {arXiv:1705.01845 [gr-qc]} \BibitemShut
  {NoStop}%
\bibitem [{\citenamefont {Necula}\ \emph {et~al.}(2012)\citenamefont {Necula},
  \citenamefont {Klimenko},\ and\ \citenamefont
  {Mitselmakher}}]{Necula:2012zz}%
  \BibitemOpen
  \bibfield  {author} {\bibinfo {author} {\bibfnamefont {V.}~\bibnamefont
  {Necula}}, \bibinfo {author} {\bibfnamefont {S.}~\bibnamefont {Klimenko}}, \
  and\ \bibinfo {author} {\bibfnamefont {G.}~\bibnamefont {Mitselmakher}},\
  }\bibfield  {booktitle} {\emph {\bibinfo {booktitle} {{Gravitational waves.
  Numerical relativity - data analysis. Proceedings, 9th Edoardo Amaldi
  Conference, Amaldi 9, and meeting, NRDA 2011, Cardiff, UK, July 10-15,
  2011}}},\ }\href {\doibase 10.1088/1742-6596/363/1/012032} {\bibfield
  {journal} {\bibinfo  {journal} {J. Phys. Conf. Ser.}\ }\textbf {\bibinfo
  {volume} {363}},\ \bibinfo {pages} {012032} (\bibinfo {year}
  {2012})}\BibitemShut {NoStop}%
\bibitem [{\citenamefont {Tiwari}\ \emph {et~al.}(2016)\citenamefont {Tiwari}
  \emph {et~al.}}]{Tiwari:2015gal}%
  \BibitemOpen
  \bibfield  {author} {\bibinfo {author} {\bibfnamefont {V.}~\bibnamefont
  {Tiwari}} \emph {et~al.},\ }\href {\doibase 10.1103/PhysRevD.93.043007}
  {\bibfield  {journal} {\bibinfo  {journal} {Phys. Rev.}\ }\textbf {\bibinfo
  {volume} {D93}},\ \bibinfo {pages} {043007} (\bibinfo {year} {2016})},\
  \Eprint {http://arxiv.org/abs/1511.09240} {arXiv:1511.09240 [gr-qc]}
  \BibitemShut {NoStop}%
\bibitem [{\citenamefont {Nuttall}(2018)}]{nuttall_detchar}%
  \BibitemOpen
  \bibfield  {author} {\bibinfo {author} {\bibfnamefont {L.~K.}\ \bibnamefont
  {Nuttall}},\ }\href {\doibase 10.1098/rsta.2017.0286} {\bibfield  {journal}
  {\bibinfo  {journal} {Philosophical Transactions of the Royal Society A:
  Mathematical, Physical and Engineering Sciences}\ }\textbf {\bibinfo {volume}
  {376}},\ \bibinfo {pages} {20170286} (\bibinfo {year} {2018})}\BibitemShut
  {NoStop}%
\bibitem [{\citenamefont {Abbott}\ \emph
  {et~al.}(2018{\natexlab{c}})\citenamefont {Abbott} \emph
  {et~al.}}]{TheLIGOScientific:2017lwt}%
  \BibitemOpen
  \bibfield  {author} {\bibinfo {author} {\bibfnamefont {B.~P.}\ \bibnamefont
  {Abbott}} \emph {et~al.} (\bibinfo {collaboration} {LIGO Scientific
  Collaboration, Virgo Collaboration}),\ }\href {\doibase
  10.1088/1361-6382/aaaafa} {\bibfield  {journal} {\bibinfo  {journal} {Class.
  Quant. Grav.}\ }\textbf {\bibinfo {volume} {35}},\ \bibinfo {pages} {065010}
  (\bibinfo {year} {2018}{\natexlab{c}})},\ \Eprint
  {http://arxiv.org/abs/1710.02185} {arXiv:1710.02185 [gr-qc]} \BibitemShut
  {NoStop}%
\bibitem [{\citenamefont {Biswas}\ \emph {et~al.}(2009)\citenamefont {Biswas},
  \citenamefont {Brady}, \citenamefont {Creighton},\ and\ \citenamefont
  {Fairhurst}}]{Biswas:2007ni}%
  \BibitemOpen
  \bibfield  {author} {\bibinfo {author} {\bibfnamefont {R.}~\bibnamefont
  {Biswas}}, \bibinfo {author} {\bibfnamefont {P.~R.}\ \bibnamefont {Brady}},
  \bibinfo {author} {\bibfnamefont {J.~D.~E.}\ \bibnamefont {Creighton}}, \
  and\ \bibinfo {author} {\bibfnamefont {S.}~\bibnamefont {Fairhurst}},\ }\href
  {\doibase 10.1088/0264-9381/26/17/175009, 10.1088/0264-9381/30/7/079502}
  {\bibfield  {journal} {\bibinfo  {journal} {Class. Quant. Grav.}\ }\textbf
  {\bibinfo {volume} {26}},\ \bibinfo {pages} {175009} (\bibinfo {year}
  {2009})},\ \bibinfo {note} {[Erratum: Class. Quant. Grav.30,079502(2013)]},\
  \Eprint {http://arxiv.org/abs/0710.0465} {arXiv:0710.0465 [gr-qc]}
  \BibitemShut {NoStop}%
\bibitem [{\citenamefont {Mroue}\ \emph {et~al.}(2013)\citenamefont {Mroue}
  \emph {et~al.}}]{Mroue:2013xna}%
  \BibitemOpen
  \bibfield  {author} {\bibinfo {author} {\bibfnamefont {A.~H.}\ \bibnamefont
  {Mroue}} \emph {et~al.},\ }\href {\doibase 10.1103/PhysRevLett.111.241104}
  {\bibfield  {journal} {\bibinfo  {journal} {Phys. Rev. Lett.}\ }\textbf
  {\bibinfo {volume} {111}},\ \bibinfo {pages} {241104} (\bibinfo {year}
  {2013})},\ \Eprint {http://arxiv.org/abs/1304.6077} {arXiv:1304.6077 [gr-qc]}
  \BibitemShut {NoStop}%
\bibitem [{\citenamefont {Healy}\ \emph {et~al.}(2017)\citenamefont {Healy},
  \citenamefont {Lousto}, \citenamefont {Zlochower},\ and\ \citenamefont
  {Campanelli}}]{Healy:2017psd}%
  \BibitemOpen
  \bibfield  {author} {\bibinfo {author} {\bibfnamefont {J.}~\bibnamefont
  {Healy}}, \bibinfo {author} {\bibfnamefont {C.~O.}\ \bibnamefont {Lousto}},
  \bibinfo {author} {\bibfnamefont {Y.}~\bibnamefont {Zlochower}}, \ and\
  \bibinfo {author} {\bibfnamefont {M.}~\bibnamefont {Campanelli}},\ }\href
  {\doibase 10.1088/1361-6382/aa91b1} {\bibfield  {journal} {\bibinfo
  {journal} {Class. Quant. Grav.}\ }\textbf {\bibinfo {volume} {34}},\ \bibinfo
  {pages} {224001} (\bibinfo {year} {2017})},\ \Eprint
  {http://arxiv.org/abs/1703.03423} {arXiv:1703.03423 [gr-qc]} \BibitemShut
  {NoStop}%
\bibitem [{\citenamefont {Jani}\ \emph {et~al.}(2016)\citenamefont {Jani},
  \citenamefont {Healy}, \citenamefont {Clark}, \citenamefont {London},
  \citenamefont {Laguna},\ and\ \citenamefont {Shoemaker}}]{Jani:2016wkt}%
  \BibitemOpen
  \bibfield  {author} {\bibinfo {author} {\bibfnamefont {K.}~\bibnamefont
  {Jani}}, \bibinfo {author} {\bibfnamefont {J.}~\bibnamefont {Healy}},
  \bibinfo {author} {\bibfnamefont {J.~A.}\ \bibnamefont {Clark}}, \bibinfo
  {author} {\bibfnamefont {L.}~\bibnamefont {London}}, \bibinfo {author}
  {\bibfnamefont {P.}~\bibnamefont {Laguna}}, \ and\ \bibinfo {author}
  {\bibfnamefont {D.}~\bibnamefont {Shoemaker}},\ }\href {\doibase
  10.1088/0264-9381/33/20/204001} {\bibfield  {journal} {\bibinfo  {journal}
  {Class. Quant. Grav.}\ }\textbf {\bibinfo {volume} {33}},\ \bibinfo {pages}
  {204001} (\bibinfo {year} {2016})},\ \Eprint
  {http://arxiv.org/abs/1605.03204} {arXiv:1605.03204 [gr-qc]} \BibitemShut
  {NoStop}%
\bibitem [{\citenamefont {Ade}\ \emph {et~al.}(2016)\citenamefont {Ade} \emph
  {et~al.}}]{Ade:2015xua}%
  \BibitemOpen
  \bibfield  {author} {\bibinfo {author} {\bibfnamefont {P.~A.~R.}\
  \bibnamefont {Ade}} \emph {et~al.} (\bibinfo {collaboration} {Planck}),\
  }\href {\doibase 10.1051/0004-6361/201525830} {\bibfield  {journal} {\bibinfo
   {journal} {Astron. Astrophys.}\ }\textbf {\bibinfo {volume} {594}},\
  \bibinfo {pages} {A13} (\bibinfo {year} {2016})},\ \Eprint
  {http://arxiv.org/abs/1502.01589} {arXiv:1502.01589 [astro-ph.CO]}
  \BibitemShut {NoStop}%
\bibitem [{\citenamefont {Portegies~Zwart}\ and\ \citenamefont
  {McMillan}(2000)}]{PortegiesZwart:2000}%
  \BibitemOpen
  \bibfield  {author} {\bibinfo {author} {\bibfnamefont {S.~F.}\ \bibnamefont
  {Portegies~Zwart}}\ and\ \bibinfo {author} {\bibfnamefont {S.}~\bibnamefont
  {McMillan}},\ }\href {\doibase 10.1086/312422} {\bibfield  {journal}
  {\bibinfo  {journal} {Astrophys. J.}\ }\textbf {\bibinfo {volume} {528}},\
  \bibinfo {pages} {L17} (\bibinfo {year} {2000})},\ \Eprint
  {http://arxiv.org/abs/astro-ph/9910061} {arXiv:astro-ph/9910061 [astro-ph]}
  \BibitemShut {NoStop}%
\bibitem [{\citenamefont {Abbott}\ \emph
  {et~al.}(2016{\natexlab{e}})\citenamefont {Abbott} \emph
  {et~al.}}]{Abbott:2016drs}%
  \BibitemOpen
  \bibfield  {author} {\bibinfo {author} {\bibfnamefont {B.~P.}\ \bibnamefont
  {Abbott}} \emph {et~al.} (\bibinfo {collaboration} {LIGO Scientific
  Collaboration, Virgo Collaboration}),\ }\href {\doibase
  10.3847/0067-0049/227/2/14} {\bibfield  {journal} {\bibinfo  {journal}
  {Astrophys. J. Suppl.}\ }\textbf {\bibinfo {volume} {227}},\ \bibinfo {pages}
  {14} (\bibinfo {year} {2016}{\natexlab{e}})},\ \Eprint
  {http://arxiv.org/abs/1606.03939} {arXiv:1606.03939 [astro-ph.HE]}
  \BibitemShut {NoStop}%
\bibitem [{\citenamefont {Abbott}\ \emph
  {et~al.}(2016{\natexlab{f}})\citenamefont {Abbott} \emph
  {et~al.}}]{Abbott:2016nhf}%
  \BibitemOpen
  \bibfield  {author} {\bibinfo {author} {\bibfnamefont {B.~P.}\ \bibnamefont
  {Abbott}} \emph {et~al.} (\bibinfo {collaboration} {LIGO Scientific
  Collaboration, Virgo Collaboration}),\ }\href {\doibase
  10.3847/2041-8205/833/1/L1} {\bibfield  {journal} {\bibinfo  {journal}
  {Astrophys. J.}\ }\textbf {\bibinfo {volume} {833}},\ \bibinfo {pages} {L1}
  (\bibinfo {year} {2016}{\natexlab{f}})},\ \Eprint
  {http://arxiv.org/abs/1602.03842} {arXiv:1602.03842 [astro-ph.HE]}
  \BibitemShut {NoStop}%
\bibitem [{\citenamefont {Hannam}\ \emph {et~al.}(2014)\citenamefont {Hannam},
  \citenamefont {Schmidt}, \citenamefont {Boh{\'e}}, \citenamefont {Haegel},
  \citenamefont {Husa} \emph {et~al.}}]{Hannam:2013oca}%
  \BibitemOpen
  \bibfield  {author} {\bibinfo {author} {\bibfnamefont {M.}~\bibnamefont
  {Hannam}}, \bibinfo {author} {\bibfnamefont {P.}~\bibnamefont {Schmidt}},
  \bibinfo {author} {\bibfnamefont {A.}~\bibnamefont {Boh{\'e}}}, \bibinfo
  {author} {\bibfnamefont {L.}~\bibnamefont {Haegel}}, \bibinfo {author}
  {\bibfnamefont {S.}~\bibnamefont {Husa}},  \emph {et~al.},\ }\href {\doibase
  10.1103/PhysRevLett.113.151101} {\bibfield  {journal} {\bibinfo  {journal}
  {Phys.Rev.Lett.}\ }\textbf {\bibinfo {volume} {113}},\ \bibinfo {pages}
  {151101} (\bibinfo {year} {2014})},\ \Eprint {http://arxiv.org/abs/1308.3271}
  {arXiv:1308.3271 [gr-qc]} \BibitemShut {NoStop}%
\bibitem [{\citenamefont {Cotesta}\ \emph {et~al.}(2018)\citenamefont
  {Cotesta}, \citenamefont {Buonanno}, \citenamefont {Boh{\'e}}, \citenamefont
  {Taracchini}, \citenamefont {Hinder},\ and\ \citenamefont
  {Ossokine}}]{Cotesta:2018fcv}%
  \BibitemOpen
  \bibfield  {author} {\bibinfo {author} {\bibfnamefont {R.}~\bibnamefont
  {Cotesta}}, \bibinfo {author} {\bibfnamefont {A.}~\bibnamefont {Buonanno}},
  \bibinfo {author} {\bibfnamefont {A.}~\bibnamefont {Boh{\'e}}}, \bibinfo
  {author} {\bibfnamefont {A.}~\bibnamefont {Taracchini}}, \bibinfo {author}
  {\bibfnamefont {I.}~\bibnamefont {Hinder}}, \ and\ \bibinfo {author}
  {\bibfnamefont {S.}~\bibnamefont {Ossokine}},\ }\href {\doibase
  10.1103/PhysRevD.98.084028} {\bibfield  {journal} {\bibinfo  {journal} {Phys.
  Rev.}\ }\textbf {\bibinfo {volume} {D98}},\ \bibinfo {pages} {084028}
  (\bibinfo {year} {2018})},\ \Eprint {http://arxiv.org/abs/1803.10701}
  {arXiv:1803.10701 [gr-qc]} \BibitemShut {NoStop}%
\bibitem [{\citenamefont {Lange}\ \emph {et~al.}(2017)\citenamefont {Lange}
  \emph {et~al.}}]{Lange:2017wki}%
  \BibitemOpen
  \bibfield  {author} {\bibinfo {author} {\bibfnamefont {J.}~\bibnamefont
  {Lange}} \emph {et~al.},\ }\href {\doibase 10.1103/PhysRevD.96.104041}
  {\bibfield  {journal} {\bibinfo  {journal} {Phys. Rev.}\ }\textbf {\bibinfo
  {volume} {D96}},\ \bibinfo {pages} {104041} (\bibinfo {year} {2017})},\
  \Eprint {http://arxiv.org/abs/1705.09833} {arXiv:1705.09833 [gr-qc]}
  \BibitemShut {NoStop}%
\bibitem [{\citenamefont {{Lange}}\ \emph {et~al.}(2018)\citenamefont
  {{Lange}}, \citenamefont {{O'Shaughnessy}},\ and\ \citenamefont
  {{Rizzo}}}]{2018arXiv180510457L}%
  \BibitemOpen
  \bibfield  {author} {\bibinfo {author} {\bibfnamefont {J.}~\bibnamefont
  {{Lange}}}, \bibinfo {author} {\bibfnamefont {R.}~\bibnamefont
  {{O'Shaughnessy}}}, \ and\ \bibinfo {author} {\bibfnamefont {M.}~\bibnamefont
  {{Rizzo}}},\ }\href@noop {} {\bibfield  {journal} {\bibinfo  {journal} {arXiv
  e-prints}\ } (\bibinfo {year} {2018})},\ \Eprint
  {http://arxiv.org/abs/1805.10457} {arXiv:1805.10457 [gr-qc]} \BibitemShut
  {NoStop}%
\bibitem [{\citenamefont {Cornish}\ and\ \citenamefont
  {Littenberg}(2015)}]{Cornish:2014kda}%
  \BibitemOpen
  \bibfield  {author} {\bibinfo {author} {\bibfnamefont {N.~J.}\ \bibnamefont
  {Cornish}}\ and\ \bibinfo {author} {\bibfnamefont {T.~B.}\ \bibnamefont
  {Littenberg}},\ }\href {\doibase 10.1088/0264-9381/32/13/135012} {\bibfield
  {journal} {\bibinfo  {journal} {Class. Quant. Grav.}\ }\textbf {\bibinfo
  {volume} {32}},\ \bibinfo {pages} {135012} (\bibinfo {year} {2015})},\
  \Eprint {http://arxiv.org/abs/1410.3835} {arXiv:1410.3835 [gr-qc]}
  \BibitemShut {NoStop}%
\bibitem [{\citenamefont {{LIGO Scientific Collaboration and Virgo
  Collaboration}}(2018)}]{bayeswave}%
  \BibitemOpen
  \bibfield  {author} {\bibinfo {author} {\bibnamefont {{LIGO Scientific
  Collaboration and Virgo Collaboration}}},\ }\href
  {https://git.ligo.org/lscsoft/bayeswave} {\enquote {\bibinfo {title}
  {{BayesWave}, https://git.ligo.org/lscsoft/bayeswave},}\ } (\bibinfo {year}
  {2018})\BibitemShut {NoStop}%
\end{thebibliography}%

\end{document}